\documentclass[12pt]{article}

\usepackage{scicite}

\usepackage{times}

\usepackage{amsmath}

\usepackage{braket}
\usepackage{bbm}
\usepackage{amsfonts}
\usepackage{graphicx}
\usepackage{listings}

\usepackage{bm}
\usepackage[colorlinks, linkcolor=blue, anchorcolor=blue, citecolor=blue]{hyperref} 
\usepackage{hyperref}
\hypersetup{hypertex=true,
	colorlinks=true,
	linkcolor=blue,
	anchorcolor=blue,
	citecolor=blue}
	
\newenvironment{sciabstract}{%
\begin{quote} \bf}
{\end{quote}}

\title{Continuous field tracking with machine learning and steady state spin squeezing}

\author
{Junlei Duan$^{1\dag}$, Zhiwei Hu$^{1}$, Xingda Lu$^{1}$, Liantuan Xiao$^{2,3}$, Suotang Jia$^{2,3}$ \\
Klaus M\o lmer$^{4\ddag}$, Yanhong Xiao$^{2,3,1\ast}$\\
\\
\normalsize{$^{1}$Department of Physics, State Key Laboratory of Surface Physics}\\
\normalsize{and Key Laboratory of Micro and Nano Photonic Structures}\\
\normalsize{(Ministry of Education), Fudan University, Shanghai 200433, China}\\
\normalsize{$^{2}$State Key Laboratory of Quantum Optics and Quantum Optics Devices,}\\
\normalsize{Institute of Laser Spectroscopy, Shanxi University, Taiyuan 030006, China}\\
\normalsize{$^{3}$Collaborative Innovation Center of Extreme Optics, Shanxi University,}\\
\normalsize{Taiyuan 030006, China}\\
\normalsize{$^{4}$Niels Bohr Institute, University of Copenhagen, Blegdamsvej 17,}\\
\normalsize{DK 2100 Copenhagen, Denmark}\\
\\
\normalsize{$^\dag$  $^\ddag$ $^\ast$ To whom correspondence should be addressed; E-mail: $^{\dag}$ 20110190074@fudan.edu.cn,}\\
\normalsize{$^{\ddag}$klaus.molmer@nbi.ku.dk,$^{\ast}$yxiao@fudan.edu.cn}
}

\date{}

\begin{document}


\baselineskip24pt


\maketitle


\begin{sciabstract}
Entanglement plays a crucial role in proposals for quantum metrology, yet demonstrating quantum enhancement in sensing with sustained spin entanglement remains a challenging endeavor. Here, we combine optical pumping and continuous quantum nondemolition measurements to achieve a sustained spin squeezed state with $\bm{4 \times 10^{10}}$ hot atoms. A metrologically relevant steady state squeezing of $\bm{-3.23 \pm 0.24}$ dB using prediction and retrodiction is maintained for about one day. We employ the system to track different types of continuous time-fluctuating magnetic fields, where we construct deep learning models to decode the measurement records from the optical signals. Quantum enhancement due to the steady spin squeezing is verified in our atomic magnetometer. These results represent important progress towards applying long-lived quantum entanglement resources in realistic settings.
\end{sciabstract}

\section*{Introduction}
Quantum enhanced metrology using entangled spins represents one of the frontiers in quantum technologies \cite{pezze2018quantum} and holds the promise to overcome the standard quantum limit (SQL) set by the spin projection noise. Spin squeezed states (SSS), where the quantum fluctuations of a certain collective spin quadrature is below that of a coherent spin state, are among the most researched entangled states of atoms. Their metrological advantage has been demonstrated in a variety of precise measurement devices ranging from atomic clocks \cite{pedrozo2020entanglement,robinson2024direct} to atom interferometers\cite{greve2022entanglement,malia2022distributed} and atomic magnetometers\cite{sewell2012magnetic,bao2020spin,zheng2023entanglement}. These quantum enhanced measurements typically involve separate preparation, evolution and detection stages, which restrict their applicability to limited time intervals within the lifetime of an initially prepared SSS.  

Tracking of continuous time-dependent signals with quantum enhancement has been demonstrated with a continuous beam of squeezed light in gravitational-wave detectors \cite{aasi2013enhanced}, optical phase tracking \cite{yonezawa2012quantum}, microscopy \cite{casacio2021quantum} and optical magnetometers \cite{wolfgramm2010squeezed,li2018quantum}. The creation of sustained spin entanglement in the presence of decoherence mechanisms has been demonstrated in trapped ions \cite{ironbit1}, superconducting qubits \cite{scbit1}, large atomic ensembles \cite{atom1} and macroscopic oscillators\cite{mechanical1,mechanical2}, but this resource has not yet, however, been used for continuous sensing. Simultaneous sensing and spin entanglement generation remains a challenge due to their mutual intervention \cite{haine2020machine}.

In addition to the preparation and the maintenance of the entangled spin state, the inference of the time-dependent strength of the external signal perturbation from the measurement record is a formidable task, due to the randomness of quantum measurements and the associated back action on the quantum state of the probe. Several inference methods are proposed in continuously monitored systems, such as maximum likelihood estimation\cite{gammelmark2013bayesian,genoni2017cramer,orenes2022improving}, Bayesian parameter estimation \cite{bouten2007introduction,belavkin1995quantum,zhang2020estimating}, and artificial intelligence (AI) based parameter estimation \cite{khanahmadi2021time}. Due to various non-ideal experimental conditions, such as electromagnetic noises, finite detector (filtering) bandwidths, random atomic motions, etc, a complete and accurate model of the system dynamics is often practically infeasible \cite{warszawski2002quantum,khanahmadi2021time} unless simplifying assumptions, such as restrictions to Gaussian signals and Gaussian states apply. Deep learning (DL) \cite{schmidhuber2015deep}, as a branch of AI, has shown capacity to learn from large amounts of complex data without any prior theoretical model. Recent years have witnessed rapidly rising interests in the application of DL in physics, for example, quantum error correction\cite{sivak2023real}, imaging\cite{orazbayev2020far,ness2020single}, identification of a spatial structure \cite{li2021photonic}, and optimization of experiments \cite{tranter2018multiparameter,vendeiro2022machine}. In the field of metrology with atomic sensors, DL has shown an advantage over physical models in identifying multi-frequency signals and vector atomic magnetometry\cite{chen2022neural,liu2022deep, meng2023machine}. In this article we demonstrate that DL does indeed offer a promising approach for continuous tracking of complicated time-dependent signals by sensors operated at the quantum level.

Here, we demonstrate a stable SSS in a continuously pumped and monitored atomic ensemble, and its application in quantum enhanced continuous field tracking with the aid of deep learning. The SSS is created by continuous quantum nondemolition (QND) measurements and is observed to persist in the laboratory for more than one day, limited only by experimental hardware imperfections. The achieved degree of steady state spin squeezing is $-3.23\pm 0.24$dB when conditioned upon the full measurement records and $-1.63 \pm 0.19$dB when conditioned only upon earlier measurements. The entangled atomic ensemble is used to track various time varying magnetic fields, including a random pulse, an Ornstein-Uhlenbeck process (OU), a double Ornstein-Uhlenbeck process (dOU), a white noise process and a general hidden Markov model. We establish and train DL models to decode the optical measurement records \cite{khanahmadi2021time} and estimate the magnetic field signal with high accuracy. The sensitivity of the random pulse magnetometer is $27.97~\textrm{fT}/\sqrt{\textrm{Hz}}$, exceeding the SQL; a quantum enhancement is also verified in the white noise experiment.

\section*{Model and Experiment}
\begin{figure}
	\centering
	\includegraphics[width=6in]{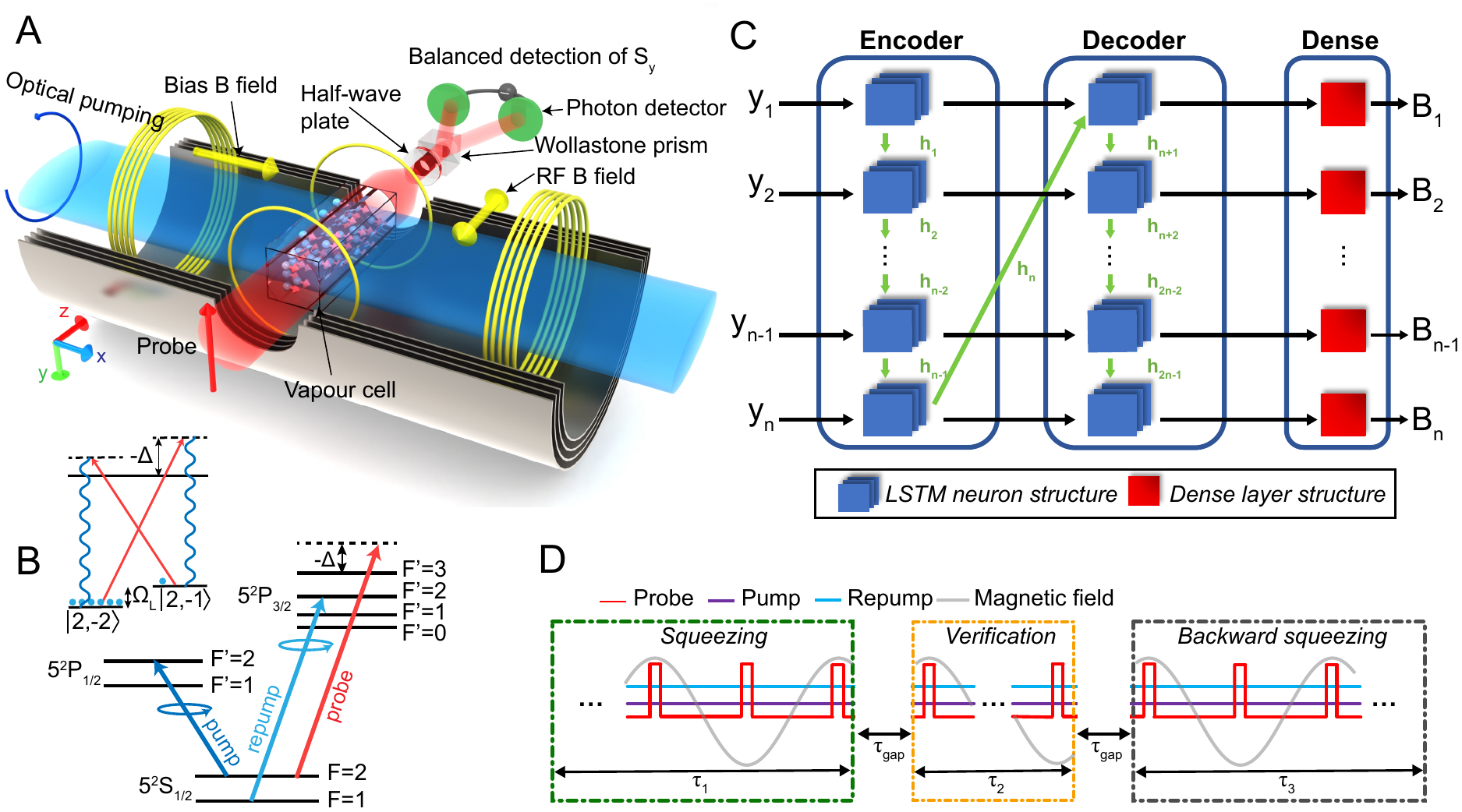}	
	\caption{{\bf Experiment schematics and working principle} 
	{\bf A.} The atomic gas is contained in a glass cell, which is heated to $55 ^\circ \textrm{C}$. The cell is coated with paraffin on its inner walls and placed in a four-layer magnetic shielding to protect the atomic spin. A bias magnetic field of $0.72~\textrm{G}$ along $x$ is applied to maintain the collective spin and leads to a sublevel Zeeman splitting of $\Omega_L = 2\pi \times 510~\textrm{kHz}$. The transmitted probe laser is detected by a balanced polarimeter and the output photocurrent is demodulated at frequency $\Omega_{L}$ by a lock-in amplifier (LIA).
	{\bf B.} Energy levels and transitions. The linearly $y$-polarized probe laser is blue detuned by $2.5~\textrm{GHz}$ to the $5S_{1/2},F=1 \rightarrow 5P_{3/2},F^{\prime}=2$ transition. 
	{\bf C.} The schematics of deep learning (DL). The DL consists of Decoder (a unidirectional LSTM layer), Encoder (a unidirectional LSTM layer) and a Dense layer. Details of the DL model can be found in SI. 
	{\bf D.} Experimental time sequence. The optical pumping lasers are continuously on and allow the system to reach a steady state. The stroboscopic intensity modulation in the probe laser at twice the Larmor frequency effectively eliminates the measurement induced quantum back-action. The average power of the probe, pump and repump is $500~\mu$W, $20~\mu$W and $400~\mu$W, respectively. The outcomes of the verification measurement sequence are correlated with the outcomes of both the prior and posterior  squeezing measurements, quantifying the level of conditional squeezing. The gap time $\tau_{gap} = 0.3~\textrm{ms}$ is set during data processing to avoid the correlation induced by the LIA.}
	\label{Fig:1}			
\end{figure}
As illustrated in Fig.~\ref{Fig:1}A, the core of the sensor is a $^{87}$Rb vapor ensemble, which consists of $N_{at} \approx 4 \times 10^{10}$ hot atoms confined in a paraffin-coated glass cell. The quantum state of the atomic ensemble can be described by collective spin operators, $\boldsymbol{\hat{J}}= \sum_k \boldsymbol{\hat{j}}^k$, where $k$ labels individual atomic spins. By continuous optical pumping, the system is retained at the coherent spin state (CSS) with mean spin $|\langle \hat{J_x} \rangle| = |J_x| = 2N_{at}$ and $\langle \hat{J}_{y,z} \rangle = 0$. The collective spin precesses around the bias magnetic field $B_{b}$ along $x$ at the Larmor frequency $\Omega_L$. The two transverse spin components $\hat{J}_{y, z}$ obey the commutation relation $[\hat{J}_{y},\hat{J}_{z}] = i \hat{J}_x$ ($\hbar = 1$), accompanied with quantum fluctuations $\Delta\hat{J}_{y} \Delta\hat{J}_{z} \geq |J_x/2| $.

%The average power of the probe is 500 $\mu$W in experiments and the associated measurement strength $\kappa^2=3000\enspace\rm{s}^{-1}$.
A spin squeezed state is generated by the measurement back action due to probing of the atoms with the off-resonant Faraday QND interaction $H_{q n d}=\left(\kappa / \sqrt{\Phi N_{a t}}\right) \hat{J}_z \hat{S}_z$, where $\boldsymbol{\hat{S}}$ is the Stokes operator of the probe light, $\Phi$ is the average photon flux and $\kappa^2$ indicates the measurement strength. The Hamiltonian evolution encodes the values of $\hat{J}_z$ on the polarization of the probe field $\hat{S}_y$, which is detected through a photon shot-noise-limited balanced homodyne polarimeter. The undesired quantum backaction (QBA) of measuring periodically varying combinations of $\hat{J}_z$ and $\hat{J}_y$ is evaded by a stroboscopic detection protocol \cite{vasilakis2015generation,bao2020spin} where the probe light intensity is modulated at twice the Larmor frequency with a duty cycle of $0.1$. We can regard the continuous measurement signal around any moment of time as composed of three sequences illustrated in Fig.~\ref{Fig:1}D: A prior interval which heralds preparation of a spin squeezed state, the current interval where verification measurements are used to validate the degree of spin squeezing, and a third, posterior measurement sequence. Spin squeezing is demonstrated by the correlation between the verification measurement outcomes and the ones obtained in the previous squeezing sequence. The verification sequence is correlated with both the outcomes of the prior measurements and of the subsequent, third measurement sequence, which demonstrates the further improvement of metrologically relevant squeezing by combined prediction and retrodiction measurements \cite{bao2020spin,gammelmark2013past,zhang2017prediction}.

% $T_1 = 48{\rm ms}$ and $T_2 = 15{\rm ms}$ respectively. In the presence of the QND probe laser, $T_1$ and $T_2$ are further reduced to $30{\rm ms}$ and $11{\rm ms}$, respectively.  
%We note that higher steady polarization is possible but at the price of harming the collective spin squeezing generation, so an optimal balance between these processes is carefully engineered in the experiment.
Due to inevitable couplings with the environment, the atomic spin experiences loss and decoherence. In our experiment, spontaneous emission, atom-atom and atom-wall collisions are dominant damping processes, which give the relaxation times for the ground state spin population and coherence, $T_1 = 30{\rm ms}$ and $T_2 = 11{\rm ms}$ respectively. In order to counter these noise processes and maintain an entangled steady state, a sustained optical pumping is applied, which drives the atoms to the $\ket{F=2, m_F=-2}$ state and hence polarizes $\hat{J}$ along the spin quantization axis ($x$). Due to the weak but resonant pumping field, $T_2$ is reduced to 2.9 ms. The degree of polarization of the atomic ensemble is measured through the magneto-optical resonance signal \cite{julsgaard2003characterizing}. Although the optical pumping itself can achieve a steady polarization (orientation) of 98.9$\%$, the presence of the probe laser decreases this value to 95.8$\%$ which gives rise to an additional 12$\%$ noise increase compared to the ideal CSS (100$\%$ polarization).

We use the setup to measure a radio-frequency (RF) magnetic field along $z$-direction, and its interaction with the atomic system can be described by a rotating-frame Hamiltonian $H_{b}(t) = 1/2 \gamma B(t) \hat{J_y}$, where $\gamma = \Omega_L/B_b$ is the gyromagnetic ratio\cite{budker_jackson2013,smith2006efficient} and $B(t)$ is the slowly varying amplitude of the RF field in a frame rotating at the Larmor frequency. The interaction induces a weak continuous rotation of the collective spin vector. Then, the evolution of the system can be described by a master equation in the rotating frame \cite{SI}.
\begin{equation}\label{eq1}
\begin{aligned}
    d \rho= -i \left[H_b(t) + H_{qnd}, \rho \right] dt +\sum_h \Gamma_h \mathcal{D}\left[c_h\right] \rho d t.
\end{aligned}
\end{equation}
Here, the commutator describes the interaction with the RF magnetic field and the QND interaction between the quantum probe field and the atoms. The second term describes the pumping and damping processes with transition  operators $c_h$ and pumping or damping rates $\Gamma_h$, and $\mathcal{D}[c_h] \rho=c_h \rho c_h^{\dagger}-1 / 2 \{c_h^{\dagger} c_h, \rho \}$. The density matrix in Eq.~\ref{eq1} represents the entangled quantum state of the atoms and the segment of the probe beam interacting with the atoms in the time interval $dt$, as governed by the input-output relation $\hat{S_y}^{(out)}(t) = \hat{S_y}^{(in)}(t) + \kappa \sqrt{S_x dt/J_x} \hat{J_z}(t)$. The field is subject to detection immediately after the interaction, and we obtain a measurement record $\mathbf{Y} = \{ Y_t \}$, where, $Y_t \sim \hat{S_y}^{(out)}(t)$.  Since the magnetic field ${\mathbf{B} = \{ B_t \}}$ drives the evolution of the monitored spin component $\hat{J_z}(t)$, we expect to learn an approximation of $\mathbf{B}$ from the record $\mathbf{Y}$. This inference is complicated because the actually measured values are subject to the quantum mechanical uncertainty of the atomic spin component $\hat{J_z}(t)$ and shot noise fluctuations and the measurements impose quantum back action on the atomic system in addition to the average evolution provided by Eq.~\ref{eq1}. Such back action makes each experimental run follow an unpredictable quantum trajectory, given by a stochastic master equation. For large systems this master equation is prohibitively complicated, and if all physical parameters and statistical properties of the noise are not fully characterized, it is not even well defined. Even though they are obviously statistically correlated, it is difficult for traditional estimation methods to establish an unbiased and reliable estimator between the measurement signal $\mathbf{Y}$ and the applied perturbation $\mathbf{B}$.

To infer the time-varying amplitude of the RF magnetic field from the optical measurements, we therefore apply a deep learning model (DL) \cite{lecun2015deep,hochreiter1997long}, whose architecture is shown in Fig.~\ref{Fig:1}C. The DL is designed with an Encoder-Decoder framework consisting of two long short-term memory layers (LSTM) and a dense layer. The input measurement records $\mathbf{Y}$ are first processed by the encoder LSTM, and the outputs are adopted as inputs to the decoder LSTM. The characteristic features of the LSTMs are that their hidden states and cell states can selectively store and forget information from previous inputs, which makes them capable of capturing the most relevant temporal correlations within the time series data. The outputs of the decoder pass through the dense layer and are mapped to the estimation $\mathbf{B_{dl}}$.

%In the experiment, the to-be-measured RF magnetic field signal $\{\mathbf{B}\}$ is produced by coils around the vapor cell, which are connected to a low noise signal generator controlled by a computer. 
The data set $\{ \{\mathbf{Y}\},\{\mathbf{B}\} \}$ is obtained by applying different magnetic fields $\mathbf{B}$ on the atomic spin and collecting the corresponding measurement records $\mathbf{Y}$ in the experiment. The data set is then divided into an independent training set and test set with a ratio of 8:2 in data volume. During the learning procedure, the DL updates its weights through a back-propagation algorithm with the Adam optimizer \cite{kingma2014adam} and the learning rate is updated by cosine annealing. The cost function is set to be the mean squared error (MSE) $\Delta^2 B = \langle (B_{real}-B_{DL})^2 \rangle$. Finally, the trained DL is able to establish a map between measurement data $\mathbf{Y}$ and a candidate applied field  $\mathbf{B}$, and the performance of the trained model is evaluated by computing the cost function on the test sets.

\section*{Results}
\begin{figure}[!h]
	\centering
	\includegraphics[width=6in]{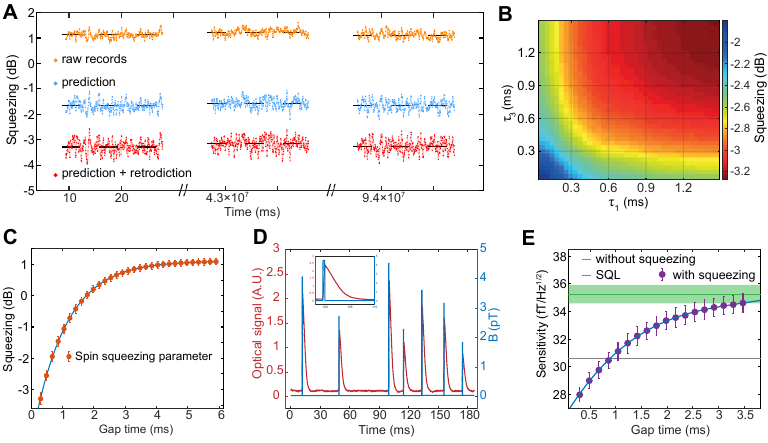}
	\caption{ {\bf Steady spin squeezing and pulse magnetometry} 
	{\bf A.} The spin squeezing $10\mathrm{lg}(\xi_W^2)$ during the total experiment time of about 26 hours. The prediction (prediction$+$retrodiction) data corresponds to  the case where the outcomes of the verification sequence are conditioned on the outcomes of the prior (prior and posterior) measurement sequences. The scattered data  comprise 500 points in time, each of which is obtained from statistical analysis of 10000 measurement records. The black dashed lines  indicate the average value.
	{\bf B.} Squeezing versus the time duration of the squeezing and backward squeezing sequence. The squeezing achieves its maximum as the sequences duration approach 1.5ms.
	{\bf C.} Squeezing versus gap time. The gap time separates the verification sequence from the prior and posterior squeezing measurements, as shown in Fig1 D. An increase of the gap time decreases the inferred spin squeezing.
	{\bf D.} The optical signal and applied magnetic field in the semi-continuous RF magnetometer. The length of a single pulse is 375~$\mu$s.
	{\bf E.} The sensitivity versus gap time in the random pulse magnetometer. The error bars are derived from 10 identical experiments, each consisting of 1000 repetitions. The blue line is to guide the eye.}
	\label{Fig:2}
\end{figure}
With the continuous optical pumping and continuous QND probing, the atomic ensemble exhibits measurement-induced steady state spin squeezing even in the presence of dissipation. In Fig.~\ref{Fig:2}A, the spin squeezing parameter $\xi_W^2$ according to the Wineland criterion \cite{wineland1994squeezed,SI} is plotted as a function of various time durations. The optical measurement data is recorded continuously for 26 hours, during which the spin squeezing is found to be maintained at $-1.63 \pm 0.19$ dB as given by the error in predicting the probe results from the outcomes of previous (QND) measurements. When applying the prediction and retrodiction QND protocol \cite{bao2020spin,gammelmark2013past,zhang2017prediction}, the probe results are confined according to a squeezing parameter of  $-3.23 \pm 0.24$ dB. The SQL is calibrated by measuring the spin noise of a completely unpolarized spin state which is insensitive to environmental noises \cite{SI}. The relationship between the squeezing and the measurement strength is shown in Fig.~\ref{Fig:2}B. It can be seen that while the optical pumping and other incoherent processes continuously decorrelate the measurements, the squeezing reaches its maximum value for measurements segments of around 1.5 ms, which is determined by the dynamical time of entanglement generation and decay.  We change the gap time between the squeezing (backward squeezing) sequence and the verification sequence in Fig.~\ref{Fig:2}C, and as expected, the squeezing level decreases for larger gap time due to decay of the entanglement.

%The reason why the signal curve in Fig.2D does not follow a perfect exponential decay is that the Larmor frequency and the carrier frequency of the magnetic field do not perfectly match in the experiment.
Having achieved steady state spin squeezing, we next examine its capability of providing a continuous metrological gain in magnetic field sensing. We first study a semi-continuous RF magnetometer whose amplitude variation is shown in Fig.~\ref{Fig:2}D. The magnetic field signal with about 180 ms total duration contains several randomly occurring fixed-width pulses with different strengths. For a measurement time of 625 $\mu$s, the standard error of magnetic field estimation enhanced by squeezing is $1.12~\textrm{pT}$, which is equivalent to a sensitivity of $27.97~\textrm{fT}/\sqrt{\textrm{Hz}}$. The above sensitivity utilizes a part of the continuous records, and can exceed the SQL as the traditional spin squeezed states do in pulsed
magnetometers \cite{bao2020spin,wasilewski2010quantum,sewell2012magnetic}. This result shows that steady spin squeezing can offer quantum enhancement for sensing the signals whenever the magnetic pulse arrives. More details of the magnetometer including the experiment setup, data processing and magnetic field calibration are outlined in supplementary materials \cite{SI}.

\begin{figure}[!h]
	\centering
	\includegraphics[width=5.1in]{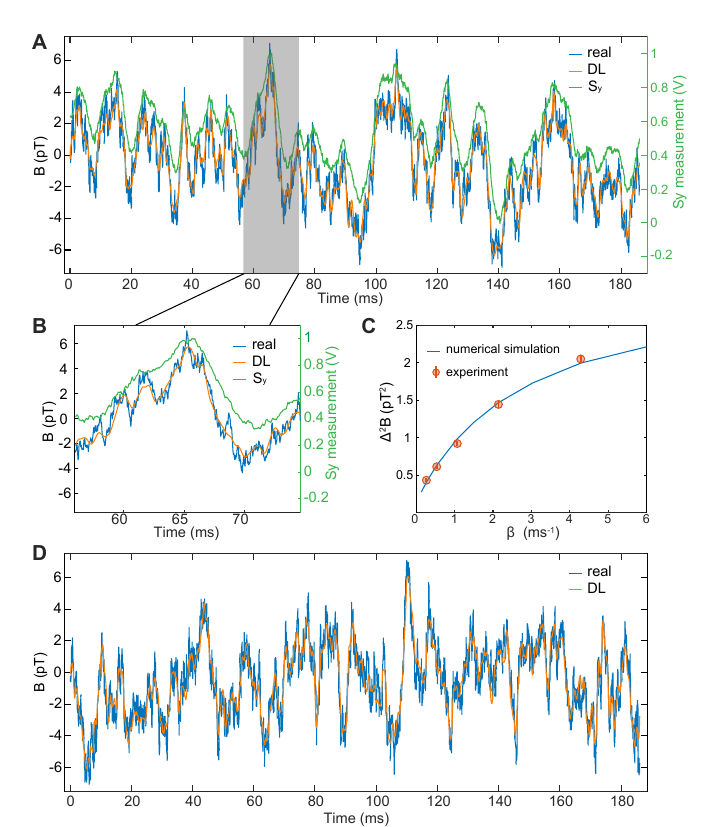}
	\noindent 
	\caption{{\bf Magnetic field tracking results} {\bf A.}  Magnetic field tracking for an OU process. For the OU process, the relaxation rate is $0.268~{ \textrm{ms}}^{-1}$ and the steady state variance is $6.12~{\textrm{pT}}^2$. The label DL indicates the output of the well-trained deep learning model. {\bf B.} A zoom-in of Fig3A in the time interval from 55 ms to 75 ms. {\bf C.} Sensitivity versus signal relaxation rate for OU signal tracking. The error bars (1 s.d.) are derived from 10 identical experiments, each consisting of 20 repetitions. {\bf D.} Magnetic field tracking for the non-Gaussian dOU process. The decay rates of the underlying OU processes are $0.402~{\textrm{ms}}^{-1}$ and $0.160~{\textrm{ms}}^{-1}$. The oscillation frequency in the weight coefficient is $\omega_d = 2 \pi \times 134~{\textrm{s}}^{-1}$ and the steady variance of the dOU process is $5.82 ~{\textrm{pT}}^2$.}
	\label{Fig:3}
\end{figure}

We then track continuous time-varying magnetic fields using deep learning models for decoding. First, we track a continuous Ornstein-Uhlenbeck (OU) process $B_{OU}$, which is described by
\begin{equation}
\begin{aligned}
    dB_{OU} (t) =-\beta B_{OU}(t) dt+ \sigma_{OU} d W_t
\end{aligned}
\end{equation}
where $d W_t$ is a stochastic Wiener increment, $\beta$ is the damping factor and $\sigma_{OU}$ denotes the noise magnitude of the OU noise process. OU processes have wide applications in finance, physics, biology, and other fields \cite{gardiner1985handbook}. OU process signals obey Gaussian statistics and they have been well studied both theoretically and experimentally \cite{yonezawa2012quantum}. Fig.~\ref{Fig:3}A shows a typical segment with a length of about $180$ ms, which contains curves for the true magnetic field, the measured probe field $S_y$ and the inferred magnetic field by the DL model. The MSE is $0.43\pm0.01~{\textrm{pT}}^2$ and is calculated from 400 repeated measurements. A noticeable feature in Fig.~\ref{Fig:3}B is that, the DL matches the overall trend of the applied magnetic field well, while small rapid changes are not tracked well, which we ascribe to the limited bandwidth of our atomic sensor. A time-normalized magnetometer sensitivity in unit of $\textrm{fT}/\sqrt{\textrm{Hz}}$ is not a meaningful quantity here as we are not estimating a constant field with accumulated precision, but a time-dependent signal with finite temporal correlations specified by the OU process parameters. The ability of the DL models to track OU processes with different correlation times is shown in Fig.~\ref{Fig:3}C. When the correlation time decreases, the field sensitivity also deteriorates. As the system dynamics are mainly described by the master equation Eq.~\ref{eq1}, we can numerically simulate the whole measurement process with different $\beta$ and calculate the corresponding MSEs with the simulated data\cite{SI}, which are in agreement with the experimental results. While traditional estimators such as Kaman filters assume Gaussian noise, the DL can be trained to track also non-Gaussian signals, and we illustrate this capability in Fig.~\ref{Fig:3}D. Here we expose the atoms to the sum of two different OU processes with weight factors that oscillate at a frequency $\omega_d$, $B_{dOU} = B_{OU1} \mathrm{cos}(\omega_d t)+B_{OU2} \mathrm{sin}(\omega_d t)$. For this non-Gaussian (dOU) process, the DL also tracks the signal well, with a MSE of $0.66 \pm 0.03~{\textrm{pT}}^2$.

\begin{figure}[!h]
	\centering
	\includegraphics[width=5.05in]{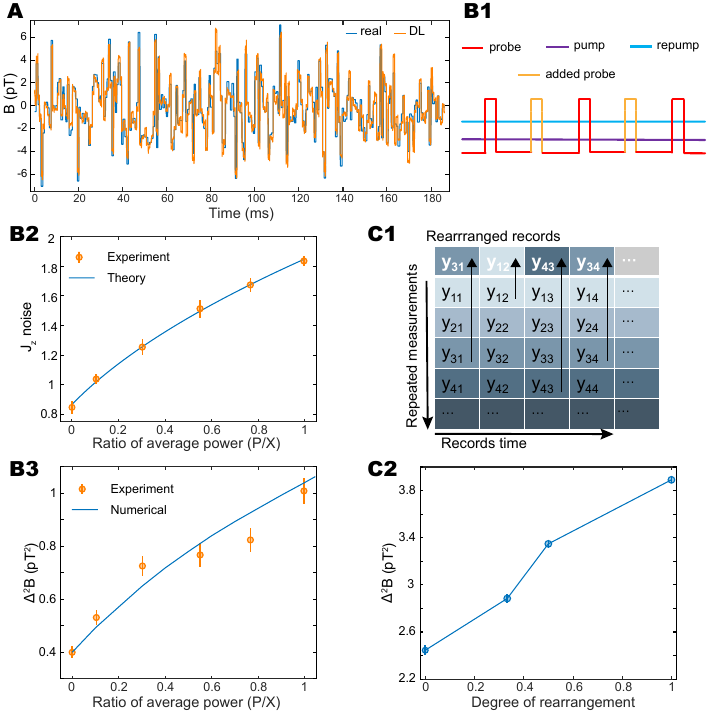}	
	\caption{{\bf Verification of quantum enhancement in continuous field tracking}  {\bf A.} Magnetic field tracking for a white noise process. {\bf B1.}The pulse sequence of the back-action experiment. Additional measurement pulses are added to probe the $J_y$ ($P$) component of the spin. {\bf B2.} The conditional noise versus the $P$ pulse strength. When the $X$ and $P$ quadratures are probed with an average power ratio equal to unity, we expect no squeezing to occur. The conditional noise is normalized to the SQL, and is well captured by the theoretical model. {\bf B3.} The magnetic field sensitivity versus $P$ pulse strength. In the rearrangement experiment, to maintain consistency with the results in QBA experiment, the polarization of the probe light is kept along $x$. We also reduce the update time increment to 373~$\mu s$ and the LIA sampling rate to eliminate the average effect of the rearrangement operation. {\bf C1.} The scheme of the rearrangement experiment. {\bf C2.} The sensitivity versus the degree of rearrangement, where 0 means no rearrangement, and 0.33, 0.5 and 1 means rearrangement of every 3, 2, 1 data points respectively. The blue line is to guide the eye.}
	\label{Fig:4}
\end{figure}

The time correlations within the OU process makes it hard to distinguish the role of atom entanglement and squeezing in the results of Fig.~\ref{Fig:3}, and to assess the quantum enhancement from SSS in continuous tracking, we turn instead to a white noise process. Fig.~\ref{Fig:4}A shows the results of white noise signal tracking. The update time increment is set to about $740~\mu$s and the MSE of the trained DL model is $0.40 \pm 0.02~{\textrm{pT}}^2$. To explicitly verify the effect of quantum back action on the collective spin noise, we perform two experiments with the schematics shown in Fig.~\ref{Fig:4}B1. We first apply additional $P$ ($J_y$) measurement pulses which introduce additional QBA and thus increase the uncertainty of $J_z$ (to eliminate the influence of higher-order terms in the light-atom interaction Hamiltonian, we changed the polarization of the probe laser from $y$ to $x$~\cite{SI}). Fig.~\ref{Fig:4}B2 shows that the $J_z$ variance (the noise is normalized to the SQL level) increases with the strengths of the $P$ pulses, as expected theoretically. Then we repeat the experiment, tracking a white noise signal with different strengths of the $P$ pulses and Fig.~\ref{Fig:4}B3 shows that the sensitivity has a similar trend as the spin noise in Fig.~\ref{Fig:4}B2. The results in Fig.~\ref{Fig:4}B2 and Fig.~\ref{Fig:4}B3 indicate that our experiment is influenced in the same way by the spin projection noise.

Next, we explore the role of the measurement induced entanglement by a rearrangement of data in the experiment. The key idea is that the quantum entanglement appears as correlations in the spin noises over the relatively short time scale of $T_2$, and we can suppress such correlations by randomly rearranging the experimental data from many repeated runs, as shown in Fig.~\ref{Fig:4}C1. Since we apply the same realization of the RF magnetic field in each repeated experiment, and the noises due to optical shot noise fluctuations are uncorrelated, any difference in the sensing capability must be due to the spin dynamics and its correlation with the actual measurement record in each run of the experiment. We compare the MSE of the DL sensing using actual recorded data with the inference based upon gradually more rearranged versions of the data in Fig.~\ref{Fig:4}C2. When we increase the degree of rearrangement, the MSE of the DL model, trained on the rearranged data, also increases. This result shows that correlations between consecutive measurements, which are due to the measurement back action on the quantum state of $J_z$ are important for the sensing scheme. Together, the effect of added QBA noise and the effect of rearranging measurement data demonstrate that the quantum measurement back action on the spins affects the sensitivity and supports that measurement induced squeezing and entanglement improves the field tracking in our experiments.

\section*{Discussion and outlook}
We have shown that DL models can be trained on experimental data and be used for sensing of fields obeying a variety of noise models. For Gaussian noise, described by an OU process, theory has shown that the DL model recovers the same sensitivity as quantum measurement theory including retrodiction \cite{khanahmadi2021time}, which clearly displays also the role of entanglement and squeezing. For non-Gaussian signals, we do not have similar reference theoretical descriptions to benchmark the DL models. While the DL models are ``theory-free", we conjecture that they, nonetheless, benefit from the measurement induced squeezing of the magnetometer. This conjecture is supported by the fact that their performance deteriorates when we alter the detection schemes to preclude squeezing. 

The degree of squeezing in our experiment is limited by the moderate optical depth and the spin decoherence, which can be improved by adopting an optical cavity \cite{vasilakis2015generation} or a longer cell\cite{zheng2023entanglement}. Applying concurrent entanglement-preparation and interrogation constitutes an interesting research topic~\cite{haine2020machine}, and may also be combined with recent insight concerning optimal quantum states for sensing with Ramsey interferometers \cite{kaubruegger2021quantum,kaubruegger2019variational}.
Our work signifies practical progress on continuous quantum sensing and the technique is applicable to other spin systems, for example, nuclear spins \cite{katz2020long,serafin2021nuclear}, diamond nitrogen-vacancy centers \cite{santagati2019magnetic,turner2022real} and mechanical oscillators \cite{rossi2019observing,meng2020mechanical}.
\\

\bibliographystyle{Science}
\bibliography{scibib}

\clearpage

\section*{Supplementary Materials}
\subsection*{S1\quad Light-atom interaction}
Our QND measurement utilizes a dispersive interaction of the probe laser with the atoms on the transition from the $5S_{1/2},F=2$ ground state to the $5P_{3/2}$ excited state. The derivation of the light-atom interaction Hamiltonian is described in \cite{julsgaard2003entanglement,hammerer2010quantum}. Due to the relatively large detuning $\Delta$, we can adiabatically eliminate the excited states and obtain the effective interaction Hamiltonian 
\begin{equation}
	\begin{aligned}
		H_{int} & =-\frac{\hbar c \Gamma \lambda^2}{8 A \Delta 2 \pi} \int_0^L\left( a_1 \hat{S}_z(z, t) \hat{\jmath}_z(z, t)\right. \\
		& \left.+a_2\left[\hat{\Phi}(z,t) \hat{\jmath}_z^2(z, t)-\hat{S}_{-}(z, t) \hat{\jmath}_{+}^2(z, t)-\hat{S}_{+}(z, t) \hat{\jmath}_{-}^2(z, t)\right]\right) \rho A {\rm d} z,
	\end{aligned}
\end{equation}
where $A$ and $L$ are the cross-section and length of the atomic medium, respectively. $\hat{\Phi}(z,t)$ is the photon flux per unit length and $c$ is the speed of light. The wavelength of the probe light is $\lambda = 780$ nm and the full width at half maximum of the atomic excited state is $\Gamma = 2 \pi \times 6.07~{\rm MHz}$. $ \hat{S}_{ \pm}=\hat{S}_y \pm i \hat{S}_z$, $ \hat{j}_{ \pm}=\hat{j}_y \pm i \hat{j}_z$ are the ladder operators of light polarization and single atomic spin, respectively.
The detuning-dependent vector and tensor polarizabilities $a_1, a_2$  are given by
\begin{equation}
	\begin{aligned}
		& a_1=\frac{\sqrt{2}}{100}\left(-\frac{15}{1-\Delta_{13} / \Delta}-\frac{25}{1-\Delta_{23} / \Delta}+140\right), \\
		& a_2=\frac{\sqrt{2}}{40}\left(\frac{1}{1-\Delta_{13} / \Delta}-\frac{5}{1-\Delta_{23} / \Delta}+4\right),
	\end{aligned}
\end{equation}
where $\Delta_{13} = 2\pi\times 423.60~{\rm MHz}$, $\Delta_{23} = 2\pi\times 266.65~{\rm MHz}$ are the hyperfine splittings in the $^{87}$Rb excited state $5P_{3/2}$. Here, $\boldsymbol{\hat{S}}$ is the optical Stokes operator. If the light propagates along the $z$-direction,
\begin{equation}
	\begin{aligned}
		\hat{S}_x & =\frac{1}{2}\left(\hat{a}_x^{\dagger} \hat{a}_x-\hat{a}_y^{\dagger} \hat{a}_y\right), \\
		\hat{S}_y & =\frac{1}{2}\left(\hat{a}_x^{\dagger} \hat{a}_y+\hat{a}_y^{\dagger} \hat{a}_x\right), \\
		\hat{S}_z & =\frac{1}{2 i}\left(\hat{a}_x^{\dagger} \hat{a}_y-\hat{a}_y^{\dagger} \hat{a}_x\right),
	\end{aligned}
\end{equation}
where the index $x (y)$ in ladder operator $\hat{a}_{x,y},\hat{a}_{x,y}^{\dagger}$ indicate the $x (y)$ polarization. In the experiment, the input $y$-polarized probe obeys $\langle \hat{S}_x \rangle = S_x = \Phi/2$ and $\langle \hat{S}_{y,z} \rangle = 0$.\\
In our vapor cell with a temperature of 55 degrees Celsius, the rubidium atoms are flying fast on the timescale of the atom-light interaction. Then we can replace the atomic spin and light Stokes operators by their averaged version
\begin{equation}
	\begin{aligned}
		\left\langle \hat{j_i}(z, t)\right\rangle_z &= \frac{1}{L} \int_0^L \hat{j_i}(z, t) d z, \\
		\left\langle \hat{S_i}(z, t)\right\rangle_z &= \frac{1}{L} \int_0^L \hat{S_i}(z, t) d z,
	\end{aligned}
\end{equation}
where $i$ labels an arbitrary component and the index $z$ indicates averaging along the $z$-direction. We can define the integrated collective atomic operators
\begin{equation}
	\hat{J}_i(t)=\int_0^L \hat{j}_i(z, t) \rho A d z.
\end{equation}
For large detuning ($\Delta = 2.5$ GHz), the terms proportional to $a_1$ is dominant ($a_2/a_1 = 0.0081$) and the $a_2$ terms can be neglected. Also, $c = \hbar = 1$ is taken for simplicity. Then we can obtain the typical QND Hamiltonian shown in the main text
\begin{equation}
	H_{q n d}= \alpha \hat{J}_z \hat{S}_z
\end{equation}
where $\alpha=-(\Gamma \lambda^2 a_1)/(16 A \Delta \pi) $, and the parameter $\kappa$ applied in the main text is given as 
\begin{equation}
	\kappa = -\frac{\Gamma \lambda^2 a_1}{16 A \Delta \pi} \sqrt{\Phi N_{at}}.
\end{equation}
In some cases, the $a_2$ terms can cause a complicated dynamical evolution and thus cannot be neglected, as explained in {\bf Section 8} of SI. 

\subsection*{S2\quad Numerical simulation}
The system evolution is directly influenced by optical pumping, damping processes, QND measurement and the to-be-measured RF magnetic field, which can be described by the master equation:\\
\begin{equation}\label{master}
	\begin{aligned}
		d \rho= - i \left[ H_b(t) + H_{qnd}, \rho \right]dt +\sum_h \Gamma_h \mathcal{D}\left[c_h\right] \rho d t,
	\end{aligned}
\end{equation}
where $H_{q n d}= \alpha \hat{J}_z \hat{S}_z$ and $H_{b}(t) = 1/2 \gamma B(t) \hat{J_y}$ indicate the effect of the QND interaction and the RF field respectively. $\mathcal{D}[c_h] \rho=c_h \rho c_h^{\dagger}-1 / 2 \{c_h^{\dagger} c_h, \rho\}$ is the Lindblad form and $\Gamma_h$ is the pumping or damping rate.  When $c_h = \hat{J_{+}}$, this term describes the pumping effect. When $c_h = \hat{J_{-}}$ and $c_h = \hat{J_x}$, this term describes the phase damping and amplitude damping process, respectively. After discretizing the continuous measurement into sequences of segments with duration $\tau$, the evolution of the spin component $\hat{J_z}$ is
\begin{equation}\label{atomevol}
	d \langle \hat{J}_z(t) \rangle = -\frac{1}{2} \gamma B(t) J_x {dt} - \Gamma_{tot} \langle \hat{J}_z(t) \rangle {dt}.
\end{equation}
Here, $\Gamma_{tot} $ is the total decay rate of $J_z$, characterized by a $T_2$ measurement. The parameter we want to estimate is $B(t)$ and the measurement records  $Y(t) \propto \hat{S_y}^{(out)}(t)$ is obtained from the balanced homodyne detection at every time segment,
\begin{equation}\label{inout}
	\hat{S_y}^{(out)}(t) =\hat{S_y}^{(in)}(t) + \alpha \sqrt{\tau} S_x \hat{J_z}(t)
\end{equation}
Eq.~\ref{inout} indicates that the atomic spin $J_z$ is continuously monitored by the light field, which could squeeze the spin degree of freedom, as shown in the next section.  The signal record $\mathbf{B} = \{B(t_0), B(t_1)...\}$ is encoded on the measurement data record $\mathbf{Y} = \{Y(t_0), Y(t_1)...\}$. We assume that the signal record and the state of the quantum systems are all Gaussian, and hence $\mathbf{Y}$ can be characterized as a conditional multi-variate Gaussian distribution
\begin{equation}\label{CMD}
	\begin{aligned}
		p(\mathbf{Y} \mid \mathbf{B})=\frac{1}{\sqrt{(2 \pi)^d\left|\boldsymbol{\Sigma}_\mathbf{B}\right|}} \exp \left[-\frac{1}{2}(\mathbf{Y}-\overline{\mathbf{Y}})^T \boldsymbol{\Sigma}_\mathbf{B}^{-1}(\mathbf{Y}-\overline{\mathbf{Y}})\right]
	\end{aligned}
\end{equation}
where $d$ is the length of $\mathbf{Y}$, $\overline{\mathbf{Y}}$ and $\boldsymbol{\Sigma}_\mathbf{B}$ are the mean vector and covariance matrix of $\mathbf{Y}$, respectively. Note that the conditional covariance matrix $\boldsymbol{\Sigma}_{\mathbf{B}}$ is different for different $\mathbf{B}$. The Cram{\'e}r-Rao bound determines the ultimate limit on the precision with which we can estimate $B_i$ with $\mathbf{Y}$ \cite{Cramer1946Mathematical},
\begin{equation}\label{CRB}
	\operatorname{Var}(B_i) \geq \frac{1}{M F(B_i)}.
\end{equation}
Here, $M$ is the number of repetitions of the measurement. The Fisher information (FI) in Eq.~\ref{CRB} for estimating $\mathbf{B}$ is
\begin{equation}
	F(\mathbf{B})=\mathbb{E} \left[\left(\frac{\partial \ln p(\mathbf{Y} \mid \mathbf{B})}{\partial \mathbf{B}}\right)^2\right].
\end{equation}
Here, $\mathbb{E}(a)$ indicates the mean value of $a$. Using the property of multi-variate gaussian distribution, the FI is
\begin{equation} \label{fi}
	F(\mathbf{B})=\left(\partial_\mathbf{B} \overline{\mathbf{Y}}\right)^{\top} \boldsymbol{\Sigma}_\mathbf{B}^{-1}\left(\partial_\mathbf{B} \overline{\mathbf{Y}}\right).
\end{equation}
Here, $\boldsymbol{\Sigma}_\mathbf{B}$ and $\partial_\mathbf{B} \overline{\mathbf{Y}}$ should be derived from the evolution of the system. However, as $\partial_\mathbf{B} \overline{\mathbf{Y}}$ and $\boldsymbol{\Sigma}_\mathbf{B}^{-1}$ are big vectors and matrices with non-trivial temporal correlations, the FI does not have a fixed analytical form but has to be obtained from solution of the equations for the mean values and the covariance matrix.

We do numerical simulations to obtain the FI and the procedure is as follows. Let us consider the experimental setup in the main text and a magnetic field that obeys the Gaussian process. Applying Eq.~\ref{master} and Eq.~\ref{inout}, we can simulate a signal measurement process for many times and obtain the magnetic field matrix $\mathcal{B}$ and the light signal matrix $\mathcal{Y}$. Elements of the matrix are labeled as $\mathcal{Y}[j,i]$ and $\mathcal{B}[j,i]$, which represent the $i$th record or magnetic field in the $j$th simulation. Then we can use the $\mathcal{Y}$ and $\mathcal{B}$ to numerically calculate Eq.~\ref{fi}. For simplicity, we consider the element $F(B_i)$, the Fisher information for the $i$th magnetic field. For the $n$th measurement record in $\mathbf{Y}$,  the element in the $\left(\partial_{B_i} \overline{\mathbf{Y}}\right)$ is
\begin{equation} \label{Der}
	\left(\partial_{B_i} \overline{\mathbf{Y}}\right)[n]=\frac{{\rm Cov}(\mathcal{Y}[:,n],\mathcal{B}[:,i])}{{\rm Var}(\mathcal{B}[:,i])},
\end{equation}
where $:$ indicate all the elements in this dimension. The covariance operation in Eq.~\ref{Der} is the covariance of two vectors, and the result is a number indicating the correlation between the corresponding elements of the two vectors. After calculating $\left(\partial_{B_i} \overline{\mathbf{Y}}\right)[n]$ for every $n$, the results contain $\left(\partial_{B_i} \overline{\mathbf{Y}}\right)$. As Eq.~\ref{CMD} is a conditional distribution, we need to first calculate the conditional covariance matrix and the inverse,
\begin{equation} \label{CovInv}
	\boldsymbol{\Sigma}_{B_i}^{-1} = {\rm Cov}\left(\mathcal{Y}- \left(\partial_{B_i} \overline{\mathbf{Y}}\right) \mathcal{B}[:,i] \right)^{-1}.
\end{equation}
Note that this covariance matrix is not conditioned on the whole magnetic field $\mathbf{B}$ but only on $B_i$. Having obtained $\left(\partial_{B_i} \overline{\mathbf{Y}}\right)$ and $\boldsymbol{\Sigma}_{B_i}^{-1}$, we can now derive the Fisher information by Eq.~\ref{fi}.
In our numerical calculations, the measurement strength $\kappa^2$, the atomic decay rate $\Gamma_{tot}$ and the signal relaxation and diffusion all take the parameters of the experiments. The length of the simulated records is 2000 and we do the simulation for $10^5$ times, leading to the evaluation of two $2000 \times 100000$ matrixes, $\mathcal{Y}$ and $\mathcal{B}$ for the calculation of the Fisher information.
The numerical simulations above only apply when the system is well characterized by Gaussian distributions of the spins and Gaussian statistics of the to-be-measured signal. 

There are many analytical and numerical results  in the literature on parameter estimation with continuously monitored quantum systems ~\cite{genoni2017cramer,zhang2020estimating,madsen2021quantum,orenes2022improving}. These include general hybrid trajectory and past quantum state theory to analyze the sensing by a single spin system of signals governed by a Hidden Markov Model (HMM), and it should be noted that sensing of non-Gaussian signals with large atomic ensembles
cannot be described by Gaussian mean values and covariances. For such cases,  we thus suggest the application of DL models to the experimental data.

\subsection*{S3\quad Entanglement generation}
In this section, we describe how continuous QND measurement and optical pumping together can enable stabilized spin squeezing, and how retrodiction (backward squeezing) works to improve the squeezing level. We first introduce the Holstein-Primakoff transformation and use the canonical operators for spin and light:
\begin{equation}
	\begin{aligned}
		\hat{x}_{\mathrm{A}}=\hat{J}_y / \sqrt{\left|\left\langle J_x\right\rangle\right|},\hat{p}_{\mathrm{A}}=\hat{J}_z / \sqrt{\left|\left\langle J_x\right\rangle\right|},\\
		\hat{x}_{\mathrm{L}}=\hat{S}_y / \sqrt{\left|\left\langle S_x\right\rangle\right|},\hat{p}_{\mathrm{L}}=\hat{S}_z / \sqrt{\left|\left\langle S_x\right\rangle\right|}.
	\end{aligned}
\end{equation}
The canonical operators obey $[\hat{x}_{\mathrm{A}},\hat{p}_{\mathrm{A}}] = i$ and $[\hat{x}_{\mathrm{L}},\hat{p}_{\mathrm{L}}] = i$. In the absence of the RF magnetic field, the system evolution is influenced by optical pumping, damping process and the QND measurement. We can employ the master equation to describe the system evolution, tracing out the light field and take the measurement outcome into account \cite{wiseman2010quantum,zhang2017prediction}:  
\begin{equation}\label{masterback}
	\begin{aligned}
		d \rho = \sum_h \Gamma_h \mathcal{D}\left[c_h\right] \rho d t + \frac{\kappa \sqrt{\eta}}{\sqrt{2}} \mathcal{H} [\hat{p}_A] \rho dW(t),
	\end{aligned}
\end{equation}
where $\eta$ is the detection efficiency and the measurement superoperator is $\mathcal{H} [c] = c \rho - \rho c^\dagger - \mathrm{Tr}\left( \rho (c+c^\dagger)\right)$. The last term describes the quantum back-action and conditional feedback through the measurement results with the mean value proportional to $\operatorname{Tr}\left(\left(J_z+J_z^{\dagger}\right) \rho\right)$ and $W(t)$ is a Wiener process. We can then derive the evolution of the variance of $p_A$:
\begin{equation}
	\frac{d V_P(t)}{dt}= - 2\Gamma_{tot} V_P(t)+ 2\Gamma_{tot} V_0 - 2 \kappa^2 \eta V_P^2(t).
\end{equation}
where $V_0$ is the measured variance of $J_z$ in the experiment. The steady state solution is
\begin{equation}\label{reducedV}
	V_P = \frac{\sqrt{1+4 V_{0} \kappa^2 \eta/ \Gamma_{tot}}-1}{2 \kappa^2 \eta/ \Gamma_{tot}}.
\end{equation}
In our experiment, $\kappa^2 = 3000\enspace\rm{s}^{-1}$, $\Gamma_{tot} = 345 \enspace\rm{s}^{-1}$ and $V_0 = 0.60$. If we assume perfect detection efficiency $\eta = 1$, the steady second-order moment of $J_z$ is found to be $V_{steady}= 0.211$, indicating $-$3.7 dB squeezing theoretically. The difference between the theoretical and the experimental squeezing levels may come from the contrast loss, additional noises, limited quantum efficiency in detection, and the fact that in characterizing experimental spin squeezing we discarded measurement records with the strongest correlation (the data during the gap time) to avoid the influence of the LIA integration process.

As we continuously monitor the system, the later measurements also carry some information about the current state just as the prior measurements. This is the essence of past quantum state protocol that makes the estimation of the current system state based on the full measurement records \cite{tsang2009time,gammelmark2013past,zhang2017prediction}. The effect of future measurement results could be described by the effect matrix $E$ acting on the system. $E$ obeys a similar stochastic master equation with an initial state $\mathbbm{I}$. The backward evolution of the system is \cite{gammelmark2013past,zhang2017prediction}
\begin{equation}
	\begin{aligned}
		d E = \sum_h \Gamma_h \mathcal{D}^{\dagger}\left[c_h\right] E d t + \frac{\kappa \sqrt{\eta}}{\sqrt{2}} \mathcal{H} [\hat{p}_A^\dagger] E dW_E(t),
	\end{aligned}
\end{equation}
where $W_E(t)$ is a Wiener process. Then the diagonal element corresponding to the second moment of $J_z$ back evolves as
\begin{equation}
	\frac{d V_R(t)}{dt}= 2\Gamma_{tot} V_R(t) + 2\Gamma_{tot} V_0 - 2 \kappa^2 \eta V_R^2(t),
\end{equation}
reaching the steady state value
\begin{equation}
	V_R = \frac{\sqrt{1+4 V_{0} \kappa^2 \eta/ \Gamma_{tot}} + 1}{2 \kappa^2 \eta / \Gamma_{tot}}.
\end{equation}
Having diagonal covariance matrices for  $\rho$ and $E$, we can determine the probability distribution for the outcome of a projective measurement of $J_z$ at time $t$ conditioned on the full measurement records. Using the Gaussian characteristics of the system, we obtain the conditional variance \cite{zhang2017prediction,zhang2020estimating}
\begin{equation}
	V_{PR}^{-1} = V_P^{-1} + V_R^{-1}.
\end{equation}
With our experimental parameters, we find the conditional variance of  $J_z$ to be $V_{steady}= 0.128$ corresponding to $-$5.9 dB ideal squeezing.

\subsection*{S4\quad Back-action of measuring $J_y$}
In this section, we discuss the impact of the additional pulses measuring $J_y$ in the white-noise magnetic field tracking experiments. In our normal signal-tracking we use stroboscopic $J_z$ measurements to avoid quantum back-action noises on $J_z$, but here we purposefully introduce back-action and counteract the squeezing by interspersing the $J_z$ measurements by $J_y$ measurements. We do this to examine if our normal experiments benefit from quantum enhancement due to squeezing. The master equation of the system in this situation is\cite{doherty2012quantum,jensen2022gaussian}: 
\begin{equation}
	\begin{aligned}
		d \rho = \sum_h \Gamma_h \mathcal{D}\left[c_h\right] \rho d t + \frac{\kappa_z \sqrt{\eta}}{\sqrt{2}} \mathcal{H} [\hat{p}_A] \rho dW_z(t)+ \frac{\kappa_y \sqrt{\eta}}{\sqrt{2}} \mathcal{H} [\hat{x}_A] \rho dW_y(t),
	\end{aligned}
\end{equation}
where $\kappa_z^2$ ($\kappa_y^2$) is the measurement strength of $J_z$ ($J_y$) and $W_{z,y}$ is a Wiener process. The second moment of $J_z$ can be derived as 
\begin{equation}
	\frac{d V_P(t)}{dt}= - 2\Gamma_{tot} V(t)+ 2\Gamma_{tot} V_0 + \frac{\kappa_y^2}{2}  - 2 \kappa_z^2 \eta V(t)^2,
\end{equation}
and the steady state solution is
\begin{equation}
	V_P = \frac{\sqrt{1+4 V_{0} \kappa_z^2 \eta/ \Gamma_{tot} + \kappa_z^2 \kappa_y^2 \eta /\Gamma_{tot}^2}-1}{2 \kappa_z^2 \eta / \Gamma_{tot}}.
\end{equation}
From these results, we see that the back-action increases the prediction uncertainty of $J_z$. With equal-strength $J_y$ and $J_z$ measurements ($\kappa_y = \kappa_z$), perfect pumping ($V_0 =0.5$) and perfect quantum detection efficiency $\eta = 1$, we obtain the steady variance $V = 0.5$, a benchmark value in the ideal situation \cite{rossi2019observing,jensen2022gaussian}.

\subsection*{S5\quad Experimental details}
The 3 mm $\times$ 3 mm $\times$ 20 mm glass vapor cell has anti-relaxation paraffin coating on the inner wall, and is placed in a four-layer magnetic shield. The laser fields intensities are controlled by acousto-optic modulators, and the pulse electrical signal is generated by the FPGA Integration Modules Opal Kelly XEM7310-A75. The photo-detector module is Thorlabs PDB450A, but with the detector units replaced by higher quantum efficiency ones (Hamamatsu). The demodulation phase of the LIA is optimized by maximizing the amplitude output of the LIA.

The homogeneous static bias magnetic field is generated by eight identical coils (with their spacings carefully designed) inside the shielding, which are connected in series to a constant current source. The pump laser is tuned to the $5S_{1/2},F=2 \rightarrow 5P_{1/2},F^{\prime}=2$ transition and the repump laser is tuned to the $5S_{1/2},F=1 \rightarrow 5P_{3/2},F^{\prime}=2$ transition, as shown in Fig.~\ref{Fig:1}B. The alignment of the bias magnetic field and the pumping field is optimized using a Bell-Bloom magnetometer configuration and the orthogonality between the bias magnetic field and the $k$ vector of the probing field is also optimized to avoid the leakage of the classical $J_x$ components into the detected quantum components of $J_y$ and $J_z$.

We use the Wineland criterion to quantify the metrological improvement of spin squeezing, $\xi_W^2 = \xi^2 / C^2$, where $C$ represents the contrast change compared to an ideal CSS state. $C$ is determined by the length of the collective spin, and $\xi^2 = {\rm Var}(J_z^{sss})/{\rm Var}(J_z^{css})$ is the noise reduction comparing to an ideal CSS state. The Wineland criterion takes into account the shortening of the spin-vector and $\xi_W^2$ represent the ratio of the minimal angular resolving power of the spin squeezed state to that of the CSS. As shown in Fig.~\ref{Fig:2}A, the QND induced noise reduction (conditional noise compared to the measured $J_z$ variance of the steady CSS state) can reach $-$2.79 dB (prediction) and $-$4.39 dB (prediction$+$retrodiction) which is larger than obtained squeezing degrees. The differences come from the shortening of the spin vector (0.37dB), imperfect orientation (0.49dB), and the classical noise.

To measure the second-order moments of the observables, identical experiments are often repeated many times to obtain its statistics (typically, $10^4$ times), for instance, in the random-pulse magnetometers. In the one-day steady spin squeezing experiment, we use signal information before and after a (to be measured) target time period to obtain its statistical information. To collect magnetometer data used for DL training, we perform 2000 identical experiment but using different forms of $B(t)$ forms.\\

\subsection*{S6\quad Characterizing the atomic orientation}
\begin{figure}[!h]
	\centering
	\includegraphics[width=6in]{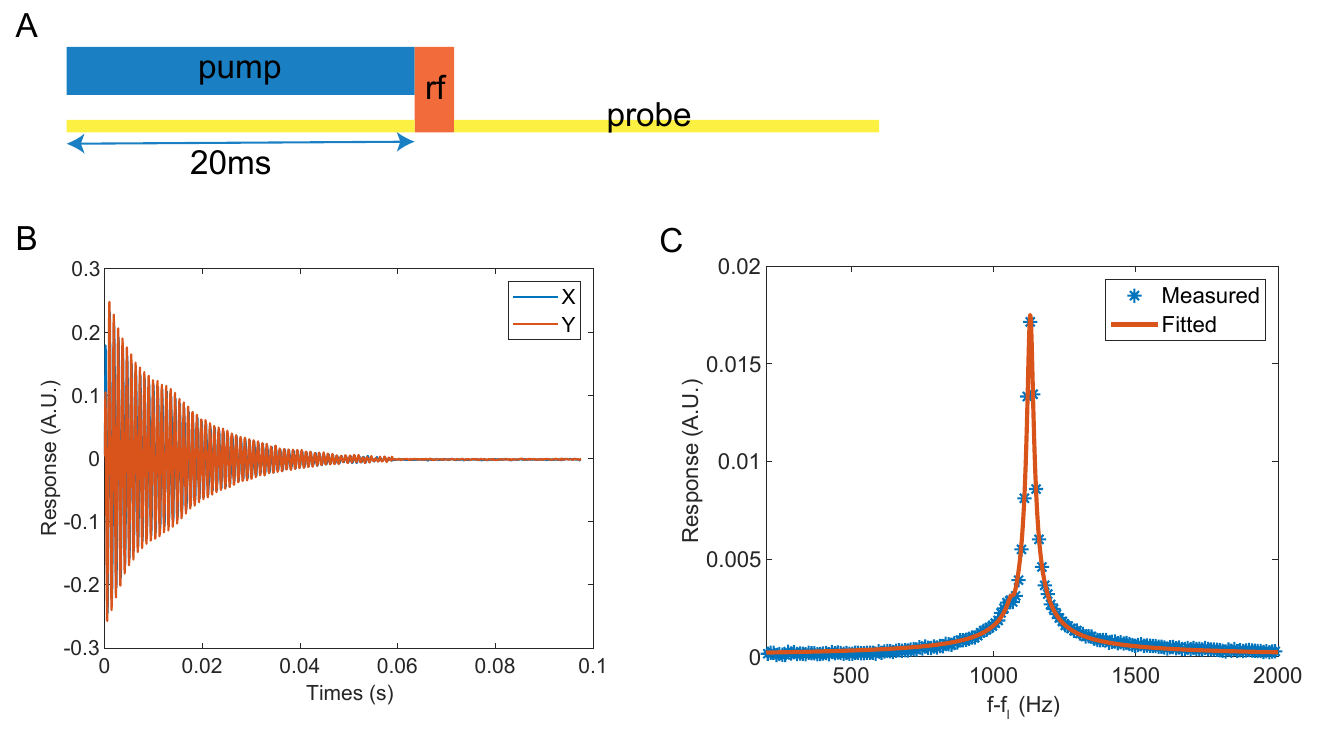}
	\caption{{\bf Magneto-optical resonance signal in the weak probe regime} A. Experimental pulse sequence. The coherent spin state is prepared by optical pumping in the presence of a weak probe field of 10${\rm \mu W}$.  B. Optically detected spin response (free induction decay) to a short RF pulse. X and Y are the two outputs of the LIA. C. Amplitude of the Fast Fourier transform of the spin response (in B) to a short RF pulse. $f_l$ is the Larmor frequency and $f$ is the actual frequency of the signal before demodulation. Fitting to the theory gives an orientation of $98.9\%$. The MORS detection power is kept at 10${\rm \mu W}$ to reduce power broadening.}
	\label{fig:S1}
\end{figure}

\begin{figure}[!h]
	\centering
	\includegraphics[width=6in]{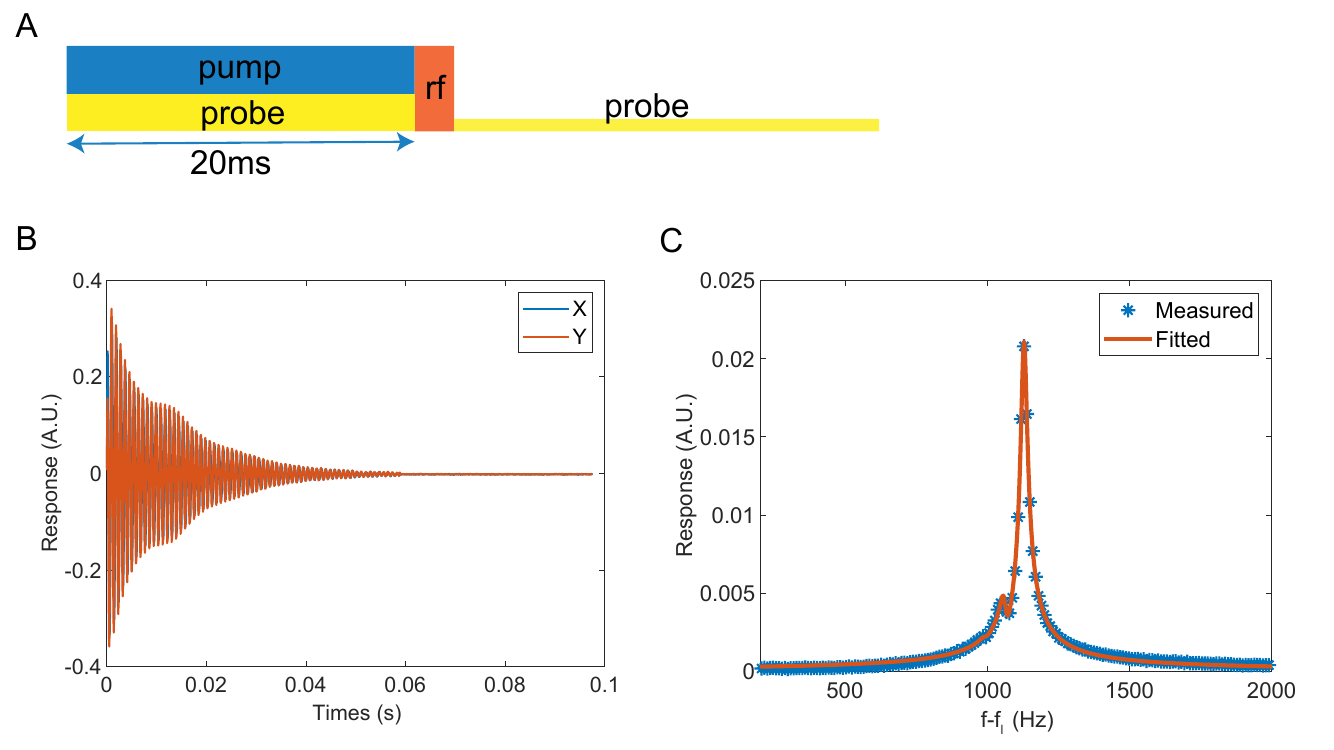}
	\caption{{\bf Magneto-optical resonance signal in the stronger probe regime} A. Experimental pulse sequence. The coherent spin state is by optical pumping but in the presence of a stronger probe field (the QND probe used in the steady state squeezing experiment) of 500$ {\rm \mu W}$.  B. Optically detected spin response (free induction decay) to a short RF pulse. X and Y are the two outputs of the LIA. It can be seen that the attenuated oscillation contains more than one frequency component. C. Amplitude of the Fast Fourier transform of the spin response (in B) to a short RF pulse.  Fitting to the theory gives an orientation of $95.9\%$, smaller than the weak probe case in Fig.S1. The MORS detection power is kept at 10${\rm \mu W}$ to reduce power broadening.}
	\label{fig:S2}
\end{figure}

\begin{figure}[!h]
	\centering
	\includegraphics[width=4in]{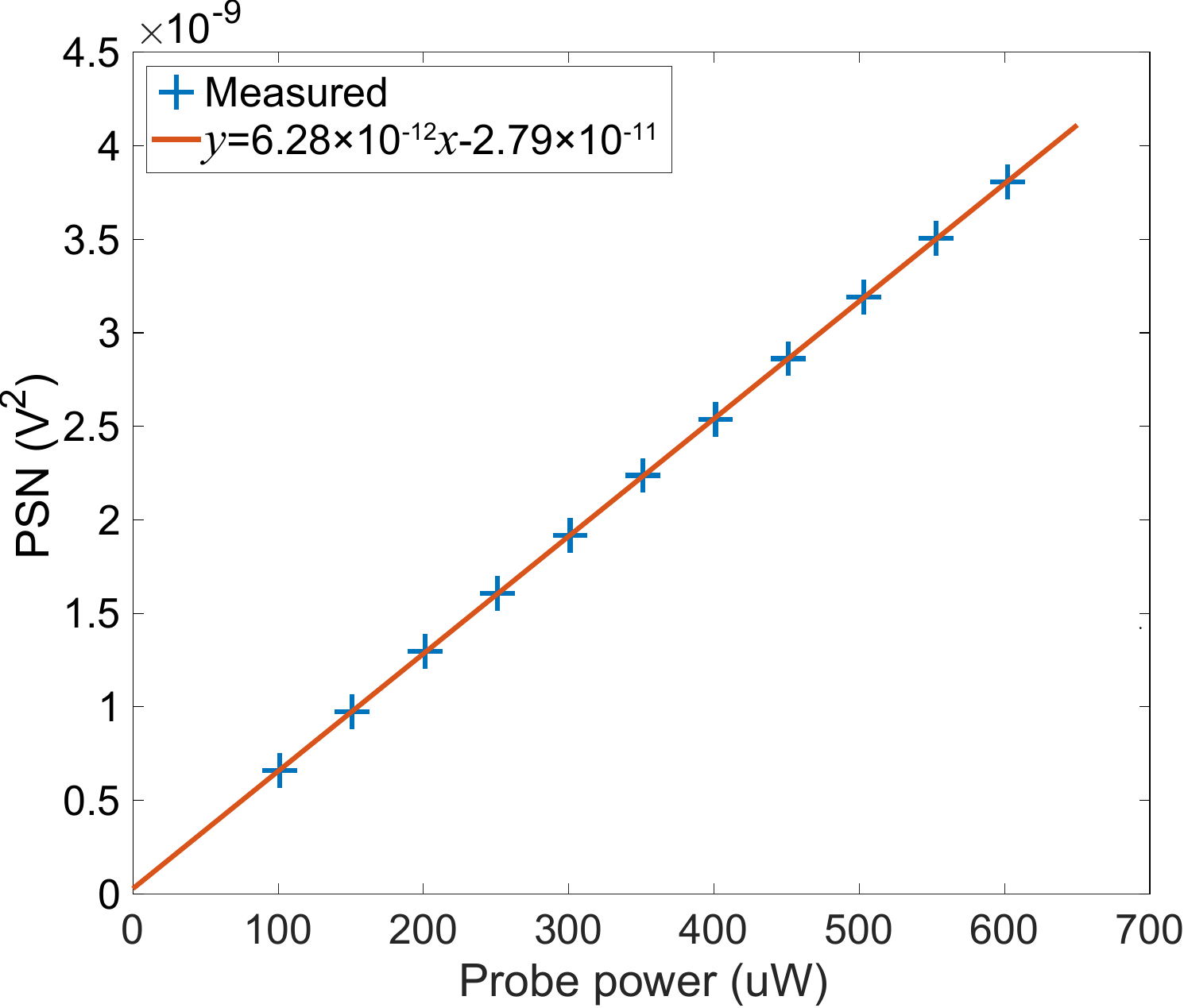}
	\caption{{\bf Photon shot noise limited optical polarimeter charaterization: the light noise versus probe power} The light noise is measured by adjusting the center frequency of atomic noise out of the response range of LIA. This is done by reducing the magnitude of the bias magnetic field. The good linearity indicates that there is little other noise in the detection system. The magnitude of measured electrical noise ($3.96\times10^{-11} \textrm{V}^2$) also matches the fitted intercept ($2.79\times10^{-11} \textrm{V}^2$). The noise is derived from $10^6$ repeated measurements.}
	\label{fig:S3}
\end{figure}

The magneto-optical resonance (MORS) method~\cite{julsgaard2003characterizing} is used to characterize the atomic orientation (i.e., the degree of the spin polarization) in the experiment. The quadratic Zeeman shift induced by the bias magnetic field makes the coherence in the magnetic sublevels of $5S_{1/2},F=2$ oscillate at four different frequencies. The measurement procedure is shown in (a) of Fig.~\ref{fig:S1} and Fig.~\ref{fig:S2}. To measure the orientation of prepared coherent spin state, first all laser fields are adiabatically turned off. Applying an RF field pulse along the $z$-direction, we then let a weak far-detuned light beam pass through the atoms along the $z$-direction. The output light is detected by a homodyne polarimeter (photon shot noise limited, as shown in Fig.~\ref{fig:S3}), and a fast Fourier transform is used to identify the different frequency components whose amplitudes are proportional to the corresponding atomic spin coherences. The population in each sublevel can then be inferred. Fig.~\ref{fig:S1} shows the case with a 10 ${\rm \mu W}$ steady probe and a spin orientation of about $98.9\%$. Fig.~\ref{fig:S2} shows the case with 500 $ {\rm \mu W}$  steady probe and the steady spin orientation is $95.8\%$, which is the condition in experiments. The MORS detection power is kept at 10 ${\rm \mu W}$ to reduce power broadening.

\subsection*{S7\quad Calibration of the standard quantum limit}
\begin{figure}[!h]
	\centering
	\includegraphics[width=4in]{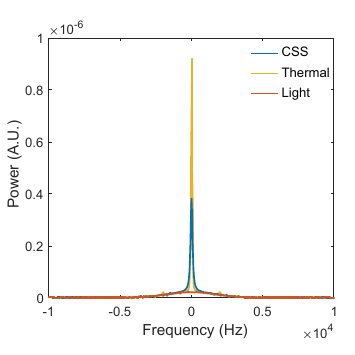}
	\caption{{\bf The noise spectrum of signals} The spectrum of CSS has two structures: the narrow sharp structure corresponds to the spin noise and has a Lorentzian shape centered at the Larmor frequency, and the broad base structure is the photon shot noise of the probe light shaped by the LIA response curve. The spin noise spectrum for the thermal state has two additional small peaks at $\pm 2kHz$, originating from the atoms on $F=1$. The difference between the CSS and thermal spin noise spectrum indicates that there are negligible contributions from the atoms in $F=1$. The linewidth difference in CSS and thermal spin noise spectrum comes from the difference in $T_2$, because the CSS noise spectrum is slightly power broadened by the continuous optical pumping beams. The light spectrum is used to obtain the response curve for LIA.}
	\label{fig:S4}
\end{figure}

The standard quantum limit is calibrated by measuring the spin noise of a completely unpolarized atomic ensemble (thermal state) and the light noise~\cite{bao2020spin,vasilakis2015generation}. The thermal state is insensitive to noises from the environmental fields or the probing laser. Meanwhile, the fractional accuracy of the noise measurement is ensured by the large number of atoms and photons used in our experiment. The thermal noise is measured when the optical pumping lasers are off, and the light noise is measured by shifting the frequency of atomic signals out of the LIA's frequency response range. Usually, we change the Larmor frequency to around 200 kHz by reducing the bias magnetic field. The standard quantum limit is
\begin{equation}
	SQL =0.8 \epsilon_{tot} \left(\operatorname{Var}\left(\hat{S}_y^{\text {thermal }}\right) - \operatorname{Var}\left(\hat{S}_y^{\text {light}}\right)\right).
\end{equation}
Here, $\operatorname{Var}\left( a \right)$ is the variance of $a$ for many experiment repetitions, and the coefficient $0.8$ comes from the population and noise difference of the thermal and coherent spin state. $\epsilon_{tot}$ is the correction coefficient accounting for experiment imperfections: (i) the CSS and thermal state have different linewidths due to optical pumping. The 3dB bandwidth of the LIA is $2$kHz, from which we derive $\epsilon_{1} = 0.98$. (ii) influences of the atom population on $5S_{1/2},F=1$ in the thermal state. As the noise from $5S_{1/2},F=1$ differs from $5S_{1/2},F=2$ by 2 kHz in the frequency domain (See Fig.~\ref{fig:S4}) due to the different Land\'e $g$-factors, we can use FFT to identify the contribution which gives $\epsilon_{2} = 0.96$. In total, we have $\epsilon_{tot} = \epsilon_{1} \epsilon_{2}$.

\subsection*{S8\quad Higher order Hamiltonian}
The terms proportional to $a_2$ is the higher-order tensor interaction. We consider the influence of this term from the Heisenberg equations of motion, which in the limit of highly-polarized atomic spin, can be approximated as 
\begin{equation}
	\begin{gathered}
		\frac{\partial}{\partial t} \hat{j}_y(z, t)=\frac{c \Gamma}{8 A \Delta} \frac{\lambda^2}{2 \pi}\left\{-a_1 \hat{S}_z \hat{j}_x+a_2\left(-\sigma^{j_x}(2 F-1)\left(2 \hat{S}_x+\hat{\Phi}\right) \hat{j}_z\right)\right\}, \\
		\frac{\partial}{\partial t} \hat{j}_z(z, t)=\frac{c \Gamma}{8 A \Delta} \frac{\lambda^2}{2 \pi} a_2\left\{4 \sigma^{j_x}(2 F-1) \hat{S}_x \hat{j}_y-2 \sigma^{j_x} \hat{j}_x\left(F-\frac{1}{2}\right) \hat{S}_y\right\}, \\
	\end{gathered}
\end{equation}
where $\sigma^{S_x}=\pm 1$ represent the $x/y$ polarization. For simplicity of the expression, we have omitted the $z,t$ arguments for the atom and light operators on the right hand side. We can then take a close look at the higher-order terms:
\begin{equation}
	\begin{aligned}
		& \frac{\partial}{\partial t} \hat{j}_y(z, t)=\alpha c j_x \hat{S}_z-\frac{\sigma^{S_x}+1}{2} \Omega_s \cdot \hat{j}_z \\
		& \frac{\partial}{\partial t} \hat{j}_z(z, t)= \sigma^{S_x} \Omega_s \cdot \hat{j}_y+\alpha c \cdot \zeta^2 \sigma^{j_x} j_x \hat{S}_y.
	\end{aligned}
\end{equation}
The definition of the parameters introduced here are $\Omega_s=\frac{c \Gamma}{8 A \Delta} \frac{\lambda^2}{2 \pi} a_2 \cdot 2(2 F-1) \sigma^{j_x} \cdot \hat{\Phi}$ and $\zeta^2 = 6a_2/a_1$. When $\sigma^{S_x} = 1$, the higher-order interaction proportional to $\Omega_s$ will be equivalent to the Larmor procession caused by a bias magnetic field and it can be well compensated by changing the stroboscopic probe frequency. When $\sigma^{S_x} = -1$, $\hat{j}_y$ will be coupled weakly and continuously to $\hat{j}_z$. The impact of this term is negligible when we only measure $J_z$.\\
When we measure both $J_z$ and $J_y$, this interaction causes $\hat{j}_z$ and $\hat{j}_y$ to convert to each other and this effect is significantly enhanced. The Heisenberg equations of motion in the case of measuring both $X$ and $P$ are
\begin{equation}
	\begin{aligned}
		& \frac{\partial}{\partial t} \hat{j}_z(z, t)= \sigma^{S_x} \Omega_s \cdot \hat{j}_y - \frac{\sigma^{S_x}+1}{2} \Omega_s \cdot \hat{j}_y +\ldots\\
		& \frac{\partial}{\partial t} \hat{j}_y(z, t)= - \frac{\sigma^{S_x}+1}{2} \Omega_s \cdot \hat{j}_z + \sigma^{S_x} \Omega_s \cdot \hat{j}_z +\ldots
	\end{aligned}
\end{equation}
where we retain only the most relevant terms for simplicity. In the case of $\sigma^{S_x} = -1$, the interaction will be non-QND if we measure the two spin components. In this situation, $\sigma^{S_x} = 1$ is a better choice to suppress the light induced non-QND interaction, which is implemented experimentally by choosing the $x$-polarization for the probe light when we use both the $X$ and $P$ measurement pulses.
\subsection*{S9\quad Data analysis}
In the main text, the squeezing level is verified by conditioning the outcome of the verification sequence on the outcome of the squeezing and backward squeezing sequences. Here, we show how this conditioning (or feedback) work. As shown in Fig.~\ref{Fig:1}D, we can divide a section of the continuous measurement $\{\mathbf{Y}\}$ into three sequences, which are squeezing, verification and backward squeezing in time order. The outcome of squeezing, verification and backward squeezing sequence could be noted as $m_1$, $m_2$ and $m_3$, respectively. Usually, $m$ is the mean value of the records in corresponding sequence. In order to fully extract squeezing, we adopted the time mode approach for $m_1$ and $m_3$. The time mode function $f(t)$ is chosen to be an exponential decay function and act as
\begin{equation}
	\begin{aligned}
		f\left(t\right) & = e^{-\Gamma_{tot}t}
	\end{aligned}
\end{equation}
\begin{equation}
	\begin{aligned}
		m_a&= \sum _t f\left(|t_0-t|\right) Y_a \left(t\right)		
	\end{aligned}
\end{equation}
where $Y_a$ is the measurement record in sequence $a$ and $t_0$ is the time mid-point for verification sequence and $\Gamma_{tot} = 1/t_2$. The time mode function weights the measurement data with time according to the decoherence rate and extracts the relevant correlations.
\begin{equation}
	\begin{aligned}
		\operatorname{Var}\left(m_2 \mid m_1\right) & =\min_\alpha\left[\operatorname{Var}\left(m_2-\alpha m_1\right)\right] \\
		& =\operatorname{Var}\left(m_2\right)-\frac{\operatorname{Cov}^2\left(m_2, m_1\right)}{\operatorname{Var}\left(m_1\right)}
	\end{aligned}
\end{equation}
Here, the feedback factor is $\alpha = \frac{\operatorname{Cov}\left(m_2, m_1\right)}{\operatorname{Var}\left(m_1\right)}$. If we apply prediction and retrodiction, the minimal variance of the measurement record $m_2$ conditioned on $m_1$ and $m_3$ is
\begin{equation}
	\begin{aligned}
		\operatorname{Var}\left(m_2 \mid m_1, m_3\right) & =\min_{\alpha,\beta}\left[\operatorname{Var}\left(m_2-\alpha m_1 - \beta m_3 \right)\right]\\
		& =\min_{\alpha,\beta} \bigr[ \operatorname{Var}\left(m_2\right) + \alpha^2    \operatorname{Var}\left(m_1\right)+ \beta^2 \operatorname{Var}\left(m_3\right) \\
		&- 2 \alpha \operatorname{Cov}\left(m_2,m_1\right)
		- 2 \beta  \operatorname{Cov}\left(m_2,m_3\right) +2 \alpha \beta \operatorname{Cov}\left(m_1,m_3\right)\bigr].
	\end{aligned}
\end{equation}
And the minimum is achieved with the feedback factors 
\begin{equation}
	\begin{aligned}
		\alpha =\frac{ \operatorname{Cov}_{2,1}\operatorname{Cov}_{3,3}-\operatorname{Cov}_{2,3}\operatorname{Cov}_{1,3}}{\operatorname{Cov}_{1,1}\operatorname{Cov}_{3,3}-\operatorname{Cov}^2_{1,3}},\\
		\beta = \frac{\operatorname{Cov}_{2,3}\operatorname{Cov}_{1,1}-\operatorname{Cov}_{2,1}\operatorname{Cov}_{1,3}}{\operatorname{Cov}_{1,1}\operatorname{Cov}_{3,3}-\operatorname{Cov}^2_{1,3}}.
	\end{aligned}
\end{equation}
Here, $\operatorname{Cov}_{a,b}$ is the covariance of $m_a$ and $m_b$.

Then, we show how the quantum-enhanced magnetometer works using the steady spin-entanglement. We consider measuring a time-varying RF magnetic field $B(t)=B_0(\omega_B) B(t)$, and $B_0(\omega_B)$ is the carrier wave form where the frequency $\omega_B$ is the atomic Larmor frequency. The amplitude modulation function $B(t)$ is to be measured, and can be derived from the spin observable $\langle J_z(t) \rangle \propto \mathcal{B}(t) = \int_{0}^{t}{B(\tau) e^{- \Gamma_{tot} |t -\tau|} d \tau}$. The probe light Stokes vector acquires a displacement from the QND interaction with the atom, whose mean value is $\langle S_y(t) \rangle \propto \langle J_z(t) \rangle$, and is detected by the optical polarimeter. The measurement records thus carry information about the unknown magnetic field. In the random-pulse magnetic field tracking, the signal consists of several pulses with random amplitudes and (non-overlapping) time of appearance. We divide the measurement records into two segments separated by the time of a pulse signal burst, say $t_1$. The measurement records after the burst time, $\{Y(t\geq t_1)\}$ gives an estimation of the field amplitude $B$ of the burst, whose sensitivity is limited by atomic projection noise and the light shot noise. As shown in Eq.~\ref{reducedV}, the atoms are in a steady entangled state, so if we incorporate the measurement records before the burst time $\{Y(t < t_1)\}$ we can obtain a sensitivity with reduced atomic projection noise. 

\subsection*{S10\quad RF magnetic field generation and calibration}
The RF field signal is produced by a signal-generator (Keysight E8257D), and its amplitude modulation signal is derived from a programmable signal-generator (Keysight 81160A). The carrier frequency is set to $2\pi \times 510$ kHz, equal to the Larmer frequency $\Omega_L$. The envelope signals are generated by a computer and sent to the Keysight 81160A through a USB connector. A pair of Helmholtz coils converts the voltage signals into a magnetic field and the atomic vapor cell is placed in the center of coils. The RF magnetic field is along the $z$-direction and this can induce a classical $J_z$ spin component in the rotating frame.

\begin{figure}[!h]
	\centering
	\includegraphics[width=6in]{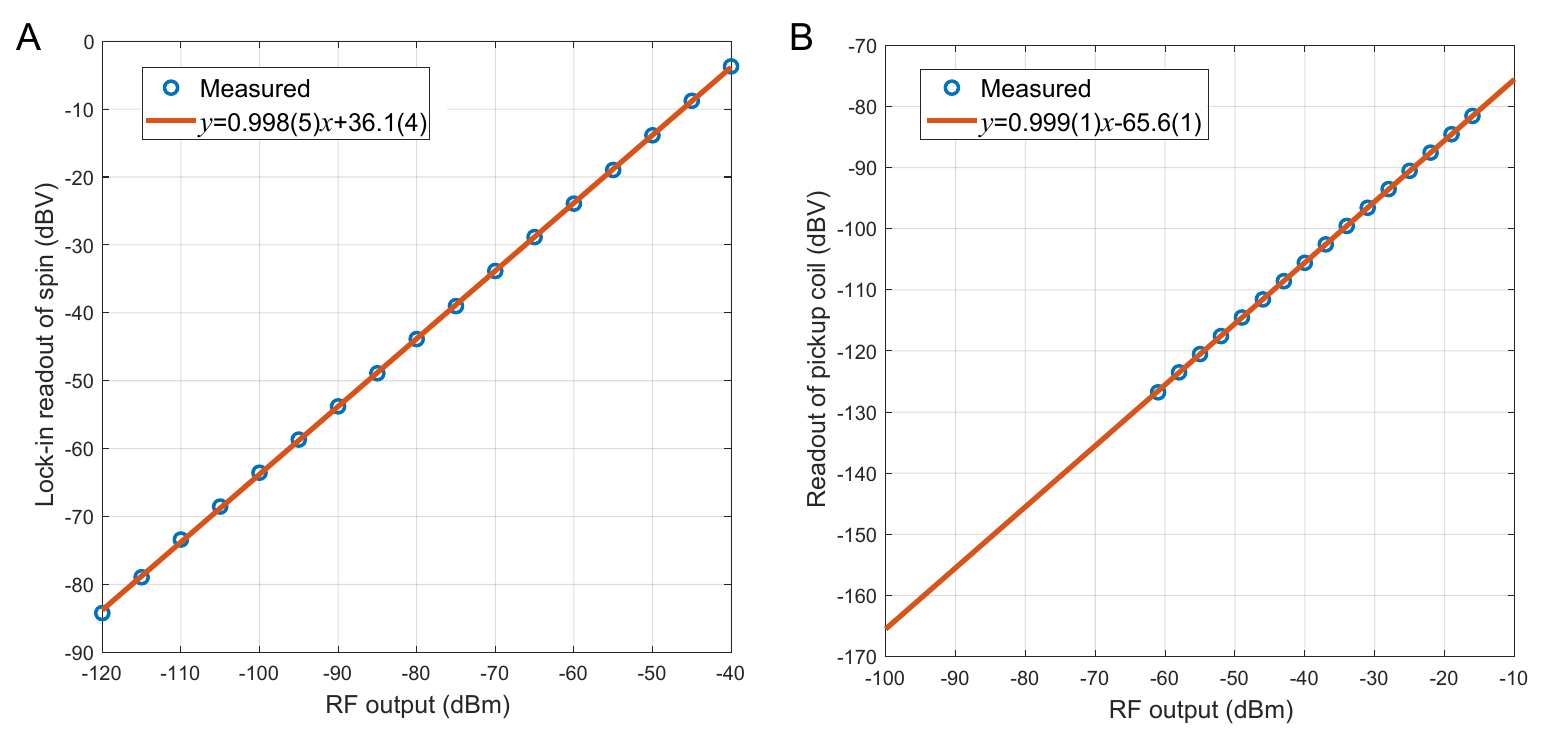}
	\caption{{\bf Calibration of the RF magnetic field} A. Calibration through the atomic response. The probe light reads the displacement of the spin and is measured with the polarimeter.  B. Calibration through the pick up coil. The RF-field induced electromotive force in the pickup coil is measured by a spectrometer. In our field-tracking experiment, the RF output from the E8257D varies from $-80 \sim -100$dBm, within the range of the calibration here. Both lines in A and B show slopes of approximately 1, indicating a good linear relation as in Eq.~\ref{cal1} and~\ref{cal2}.}
	\label{fig:S5}
\end{figure}

A pickup coil is used to calibrate the magnetic field generated by the RF coil before the magnetometer experiment. The pickup coil is placed where the Rb cell is located, and its axis is aligned along the Helmholtz coils' axis. The pickup coil thus detects the oscillating magnetic field through the generated electromotive force. Quantitatively, a sinusoidal magnetic field with frequency $\Omega_L$ and amplitude $B_{RF}$ is generated by the Helmholtz coils, and the induced electromotive force in the pickup coil is detected by a connected spectrum analyzer. The relationship between the measured signal of the spectrum analyzer $U_{SA}$ and $B_{RF}$ is \cite{bao2020spin}
\begin{equation}\label{cal1}
	\begin{aligned}
		\left|B_{\mathrm{RF}}\right|=\frac{\sqrt{2}\left|1+Z_{\text {coil }} / R_{\mathrm{m}}\right|\left|U_{SA}\right|}{N_\omega A_{\text {coil }} \omega}
	\end{aligned}
\end{equation}
The parameters of the pickup coil are shown in table \ref{pickupcoil}. The internal resistance of the spectrum analyzer is $R_{\mathrm{m}} = 50\Omega$. The results of the calibration are shown in Fig.~\ref{fig:S5}, where (a) and (b) show the calibration through the pick-up coil and atomic response, respectively. The pick-up coil calibration is effective for relatively large RF output and is not sensitive enough to reach the RF field's amplitude range used in the tracking experiment ($-$80 $\sim$ $-$100dBm). We can assume that the $B_{RF} \propto U_{E8257D}$ is valid when $U_{E8257D}$ is small, where $U_{E8257D}$ is the output voltage of the Keysight E8257D. Then we can extrapolate the line in Fig.~\ref{fig:S5}B to the left down area, which is the regime in our experiments. The validity of this assumption is verified by using the atoms to measure the $B_{RF}$ when $U_{E8257D}$ is relatively small. Atoms are more sensitive sensors and can be used to prove the linear relation between the set output power of E8257D and the magnetic field over a large range ($-$40 $\sim$ $-$120dBm). Then, we can obtain the relation between the $B_{RF}$ and the set output power of the signal generator
\begin{equation}\label{cal2}
	B_0(T)=1.428 \times 10^{-7} \times 10^{{P_{set} (dBm)} / 20}
\end{equation}
Here, $P_{set}$ is the output power we set on the signal generator. In the continuous tracking experiment, we usually set $P_{set}$ to be about $-80$ dBm and the corresponding $B_{RF}$ is about $14$ pT.

\begin{table}
	\caption{The parameters of the pickup coil}
	\begin{minipage}{0.3\linewidth}
		\centering
		\resizebox{1\textwidth}{!}{
			\begin{tabular}{lc}
				\hline
				wire turns $N_{coil}$ & 90     \\
				diameter $d_{coil}$   & 10.5mm \\
				inductance $L_{coil}$ & 60.9 $\mu$H  \\
				resistance $R_{coil}$ & 2.2 $\Omega$ \\
				impedance  $Z_{coil}$ & $i\cdot 195 \Omega$ \\
				\hline			
			\end{tabular}
		}
	\end{minipage}	
	\label{pickupcoil}
\end{table}

\subsection*{S11\quad Electromagnetic shielding}
Polarized spins are highly sensitive magnetic field sensors, which also means that good electromagnetic shielding is needed so that the background noise effect is minimized and stable. In our experiment, the ambient magnetic fields are well isolated by the four-layer magnetic shield. However, although they are located inside the magnetic shield, the coils producing the holding magnetic field and the to-be-measured RF field, may induce electromagnetic noise via their connection to lab instruments and hence greatly elevate the measured atomic spin noise far above the SQL. We have developed the following approaches to suppress such noises. \\
1. A single-phase three-stage high-performance power filter (SJD710) is used to provide a clean power supply specifically for the RF magnetic field generation system.\\
2. The cable that connects to the RF-field coil is designed with shielded twisted pairs and a short length. The shield of the cable is also well grounded.\\
3. Several resistors and inductors are added between the magnetic field generation system and the laboratory ground to obtain good grounding and minimize the influence of the ground loop. We note that consultation with the instrument manufacturer is advised to avoid endangering the equipment.\\
4. A clean source for the RF-field generation is needed and we used the Keysight E8257D model whose noise is negligibly low in our experimental frequency range. However, the noise property of the signal generator for modulating the RF-field's amplitude is less important.\\
5. The signal generators for the magnetic fields need to be connected to other equipments, where the introduced electromagnetic noise also needs to be minimized. Two common mode filters have been added, one in the time base locked loop and another in the modulation signal trigger loop. A USB digital isolator (ADUM4160) is utilized to isolate the electromagnetic noise from the computer side.\\
As a final note, the optimization methods here are all tried out via the variable-control approach and their effects may vary with the environment. The key is that our atomic system itself is a high-performance field sensor and works at a relatively high frequency, which allows us to use the atomic response as a convenient and accurate criterion in the optimization process for best electromagnetic shielding.

\subsection*{S12\quad DL model}
As shown in Fig.~\ref{Fig:1}C, the structure of the algorithm can be described as follows. Overall, the DL model we use consists of two connected unidirectional LSTM layers and one fully connected layer. For a unidirectional LSTM layer, we input the corresponding optical signals $Y_1$ to $Y_n$ at $n$ nodes arranged in a time series. For any time node $i$, the inputs consist of two parts: one is the optical signals $Y_i$ (collected from lock-in or from the previous LSTM layer), and the other is the hidden state $h_{i-1}$ transmitted from the previous time node. Both will be fed into the LSTM cells with a hidden dimension of 128. After a series of gate operations, the output results which include the signals $Y_i^{'}$ (which will be input to the next layer) and the updated hidden state $h_{i}$, are transmitted to the next time node $i+1$. The reason for arranging a continuous two-layer unidirectional LSTM structure is to extract complete information over the time series in the first layer and utilize it at each time node in the second layer. The signal $Y_i^{''}$ output by the second layer of unidirectional LSTM will be input into the fully connected layer, ultimately yielding the predicted magnetic field signal $B_i$ at each time node.

\begin{figure}[!h]
	\centering
	\includegraphics[width=6in]{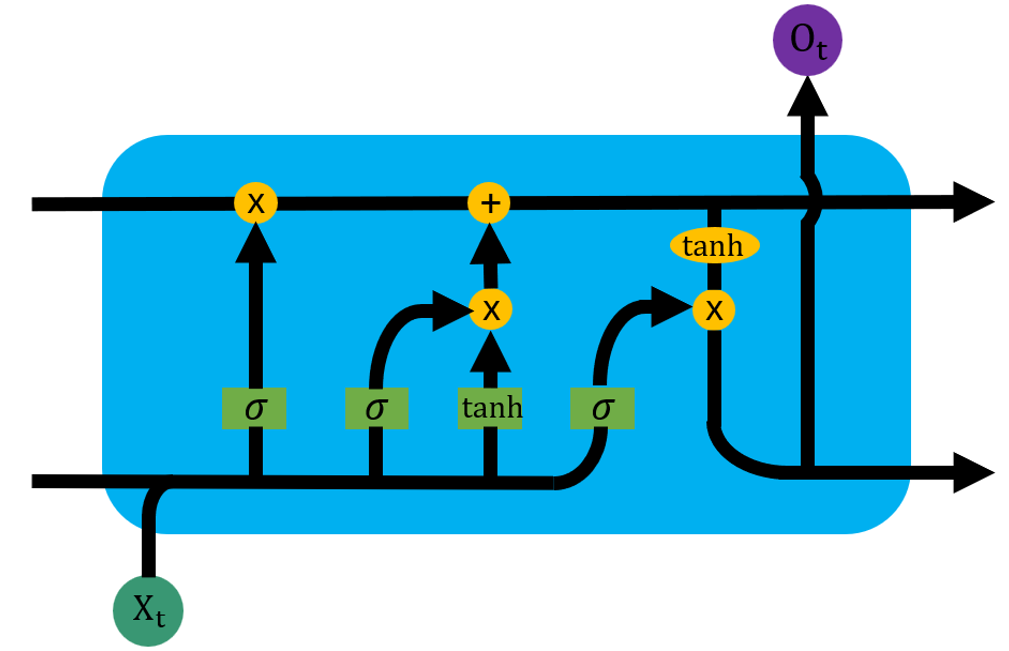}
	\caption{{\bf The cells of the LSTM layer}}
	\label{fig:S6}
\end{figure}

As a type of recurrent neural network (RNN) algorithm, LSTM has basic module units similar to, but more complex than those of the ordinary RNN algorithms. As shown in Fig.~\ref{fig:S6}, it is a basic repetitive module (also known as a neuron) of LSTM.
In the LSTM algorithm structure, cell state $C_t$ and hidden state $h_t$ are transmitted between neurons, as shown in Fig.~\ref{Fig:1}C. Both can be seen as a ``memory" for all previously input data to the recursive neural network, and they will be updated within each neuron with the new input $x_t$. The vector $C_t$ mainly records the longer term ``memory", which includes the neural network's ``preferred memory" of all previous input information. The vector $h_t$ is updated faster and it mainly records ``new memories" with shorter duration. The ``forgetting gate" determines which  previously recorded memories must be forgotten based on the new input $x_t$ and the hidden state $h_{t-1}$ of the previous moment. The ``forgotten" coefficient is passed to cell state $C_t$ to guide the change of the long-term memory. The formal expression is as follows:
\begin{equation}
	f_t=\sigma(U_f h_{t-1}+W_f x_t + b_f)
\end{equation}
where $\sigma(x) = 1/(1+e^{-x}) $ is the sigmoid function. The ``input gate" is used to control whether the newly input data $x_t$ is written into the cell state. It can filter and prioritize the input information, and higher-priority information is more likely to be written into the new cell state $c_t$. The formula is as follows:
\begin{equation}
	i_t=\sigma(U_ih_{t-1}+W_ix_t + b_i),
\end{equation}
\begin{equation}
	g_t=tanh(U_gh_{t-1}+W_gx_t + b_g),
\end{equation}
\begin{equation}
	c_t= g_t \times i_t + c_{t-1}\times f_t.
\end{equation}
The ``output gate" integrates the information in the current input value $X_t$ with the information in the previous cell state $C_{t-1}$ that has been updated through the ``forgetting gate" and ``input gate", and outputs it. The formula is as follows:
\begin{equation}
	O_t=\sigma(U_oh_{t-1}+W_ox_t+ b_o),
\end{equation}
\begin{equation}
	h_t=tanh(c_t)\times O_t.
\end{equation}

\begin{figure}[!h]
	\centering
	\includegraphics[width=6in]{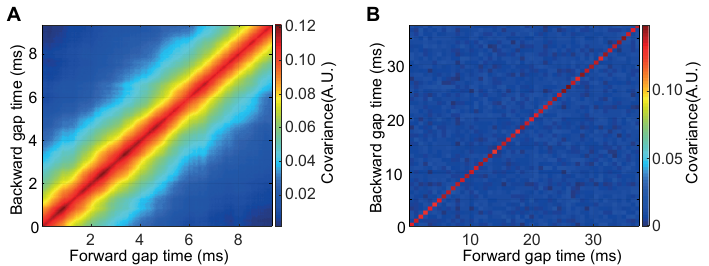}
	\caption{{\bf Covariance matrix of the applied signals} A.The OU process. The covariance shows that the signal is correlated in time.  B. The white noise. The covariance shows that the signal is not correlated in time.  The width of the red squares indicate the update time interval.}
	\label{fig:S7}
\end{figure}

\begin{figure}[!h]
	\centering
	\includegraphics[width=6in]{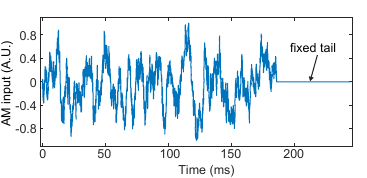}
	\caption{{\bf The fixed tail in the AM input} The signal is a dOU process. The fixed tail is used to calibrate the atomic response as background DC magnetic field slowly varies during the relatively long period of data collection.}
	\label{fig:S8}
\end{figure}

\begin{figure}[!h]
	\centering
	\includegraphics[width=6in]{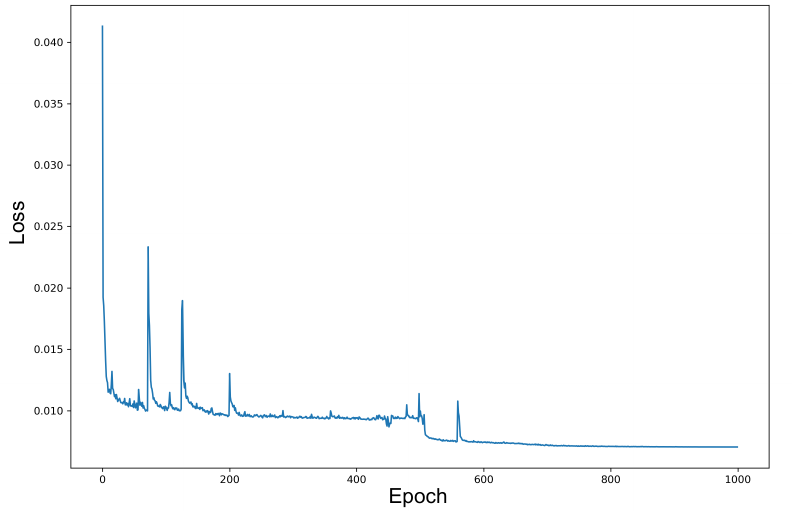}
	\caption{{\bf Loss curve} Loss curve evolution with epochs for the training set. The data size of training set is 1600, each of which contains 4986 measurement points. The loss curves for training sets converge at 800 epoch and the test sets having loss functions of similar sizes.}
	\label{fig:S9}
\end{figure}

In the experiments, we input thousands of sets of magnetic field signals which conform to specific functional relationships (such as the OU, dOU or white noise processes) into the experimental system. These sets are independent of each other (see Fig.~\ref{fig:S7}). Typically, about twelve hours are needed to obtain enough data sets. During this period, some experimental conditions may change, such as the long-term slow variation of the magnetic field background inside the magnetic shields. This can cause variations of the spin response in different data sets. To address this issue, we add a fixed tail at the end of each signal, which is a known constant magnetic field for calibration, as shown in Fig.~\ref{fig:S8}. The fixed tail allows the DL models to detect and take into account the environmental fluctuations. Each set of magnetic field signals input has 6561 signal points (4986 points for the signal, 1575 points for the fixed tail, see Fig.~\ref{fig:S8}) in a time series and we collect corresponding sets of optical signals in the detection part of the experimental system, which also has 6561 points. Before the training process, we randomly divide these sets of optical signal data (and the corresponding actual magnetic field signal data) into training and testing sets in a ratio of 8:2 for DL training. The batchsize is 40 in the training, which is limited by the GPU memory. A typical loss curve is shown in Fig.~\ref{fig:S9}.

The network is implemented using the torch 1.13.1 framework and CUDA 11.6 in Python 3.8.8. All weights are initialized with the torch default. The training process is run on a computer with CPU Intel(R) Core${\rm ^{TM}}$ i7-8700 and GPU NVIDIA GeForce RTX 2070 (8G RAM).

\subsection*{S13\quad The rearrangement experiment}
To verify quantum enhancement in the tracking of the white noise magnetic field, we perform the rearrangement experiment to identify short term correlation between the measurement records. In the experiment, every generated RF signal is applied 50 times, and the corresponding $S_y$ is also recorded 50 times, and the results are rearranged as shown in the main text.

Here, we notice that the rearrangement experiment itself is a weak proof, where the trend with rearrangement degree in Fig.~\ref{Fig:4}C2 only indicate that there exist correlations between nearby measurement records. We wish to verify that there is entanglement and that this enhances the MSE as expected for spin squeezed states. As the system noise is dominated by the quantum noise, we could make a simple assumption that the DL just subtracts the posterior and prior signals and the feedback coefficient is 1. Considering that $\kappa^2 = 3$ $\mathrm{ms^{-1}}$, the update time interval is 0.375 $\mathrm{ms}$, $LN$ ($AN$) is the light noise (atom noise) for each data point, we have $AN = 1.125 LN$.\\
a) Without rearrangement, the atomic noise is correlated.\\
The ideal MSE equals the reduced atomic noise plus light noise, $2LN + AN = 3.125LN$.\\
b) With rearrangement, the atomic noise is not correlated.\\
The noise then equals two times the atomic noise plus the light noise, $2LN + 2AN = 4.25AN$.

Here, we can define the strong standard: the MSE is lower than $3.125LN$ as the case in a) where the atomic noise is reduced due to quantum entanglement. Fig.~\ref{Fig:4}C2 shows that the MSE of the fully rearranged data is 3.89 $\rm{pT}^2$, then the stronger standard is $3.89/4.25 \cdot 3.125 = 2.86$ $\rm{pT}^2$. We can see that the MSE without rearrangement is 2.44 $\rm{pT}^2$, which is lower than the strong standard.

\subsection*{S14\quad HMM model}
\begin{figure}[!h]
	\centering
	\includegraphics[width=6in]{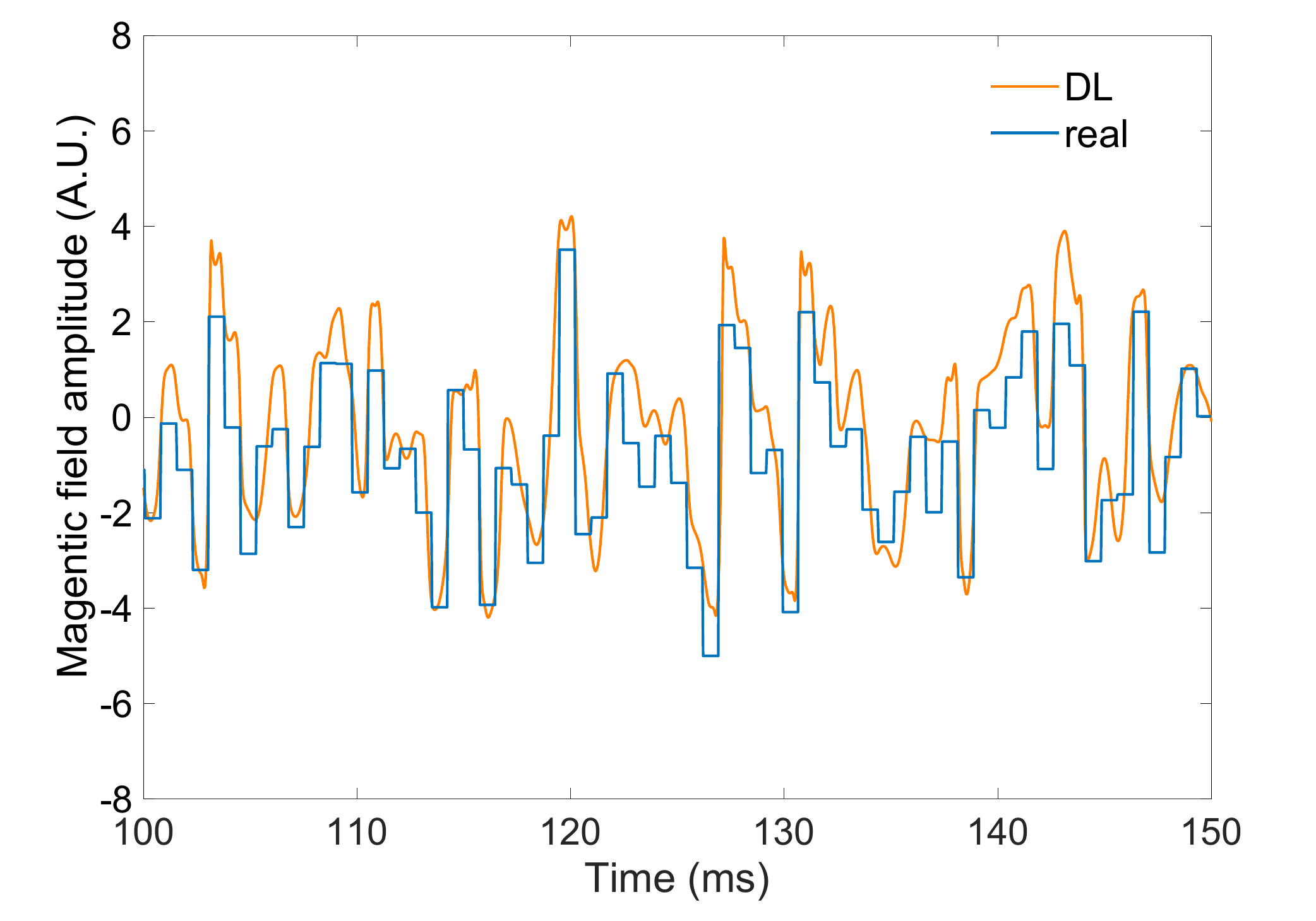}
	\caption{{\bf Signal tracking for HMM} The RF magnetic field amplitude is modulated according to a  hidden Markov model (HMM). The blue curve shows the true amplitude of the B field and the orange curve shows the output prediction of the LSTM model. The HMM model is described briefly in the text of this supplementary material.}
	\label{fig:S10}
\end{figure}

The hidden Markov models (HMMs) provide a framework for describing complex systems by sequences of observable random variables, while accounting for the hidden features of the data. Generally, a HMM consists of two random processes: a hidden state process and an observable output process. The hidden state process is a Markov chain, where each state has a probability of transitioning to one other state. The observable process is related to the hidden state process through a state dependent emission distribution. HMMs are powerful tools and have been applied in various scenarios including recognition of speech  and of handwriting, prediction of financial markets, protein folding, etc.
We track a HMM to show that our atomic sensor has the potential to track different types of signals. The HMM we use here has ten hidden states and the tracking performance is shown in Fig.~\ref{fig:S10}.
%The HMM we use here has ten hidden states and the generating function can be found in the code. The tracking performance is shown in Fig.~\ref{fig:S10}.

\clearpage

\clearpage

\clearpage

\clearpage

\clearpage

\clearpage

\clearpage

\clearpage

\clearpage

\end{document}